\documentclass[11pt,a4paper]{article}
\usepackage{amsmath,amssymb,amsthm,amsfonts}
\usepackage[dvips]{graphicx}
\usepackage{epsfig}
\usepackage{siunitx}	
\usepackage{subfigure}
\usepackage{float}
\usepackage{verbatim}
\usepackage[font=small]{caption}
\usepackage{soul}

\usepackage{color}

\title{Mathematical modelling of carbon capture in a packed column by adsorption}
\author{T.G. Myers\footnote{Centre de Recerca Matem\`{a}tica, Campus de Bellaterra  Edifici C, 08193 Bellaterra, Barcelona, Spain.}, F. Font\footnotemark[1], M. G. Hennessy\footnote{Mathematical Institute, University of Oxford, Oxford, OX2 6GG, United Kingdom} }

\def\p{\partial}

\def\({\text{\huge (}}
\def\){\text{\huge )}}

\def\]{\text{\huge ]}}
\def\[{\text{\huge [}}

\providecommand{\keywords}[1]{\textbf{\textit{Keywords:}} #1}

\newcommand{\et}{\emph{et al}.\ }
\newcommand{\bi}{\begin{itemize}}
\newcommand{\ei}{\end{itemize}}
\newcommand{\be}{\begin{equation}}
\newcommand{\ee}{\end{equation}}
\newcommand{\ba}{\begin{align}}
\newcommand{\ea}{\end{align}}

\newcommand\nc{\newcommand}
\nc\pad[2]{\frac{\p #1}{\p #2}} \nc\padd[2]{\frac{\p^2 #1}{\p
{#2}^2}} \nc\nd[2]{\frac{d #1}{d #2}} \nc\pat[2]{\frac{D #1}{D
#2}} \nc\ov{\overline} \nc\degree{^{\circ}} \nc\ord[1]{{\cal
O}(#1)} \nc\ra{\rightarrow} \nc\Ra{\Rightarrow} \nc\dint{{\mbox ~
d}}

\newcommand{\bea}{\begin{eqnarray}}
\newcommand{\eea}{\end{eqnarray}}
\newcommand{\beas}{\begin{eqnarray*}}
\newcommand{\eeas}{\end{eqnarray*}}

\begin{document}
\maketitle

\begin{abstract}
A mathematical model of the process of carbon capture in a packed column  by adsorption is developed and analysed.
First a detailed study is made of the governing equations. Due to the complexity of the internal geometry it is standard practice to average these equations. Here the averaging process is revisited. This shows that there exists a number of errors and some confusion in the standard systems studied in the literature. These errors affect the parameter estimation, with consequences when the experimental set-up is modified or scaled-up. Assuming, as a first approximation, an isothermal model the gas concentration equation is solved numerically. Excellent agreement with  data from a pressure swing adsorption experiment is demonstrated. A new analytical solution (valid away from the inlet) is obtained. This provides explicit relations for quantities such as the amount of adsorbed gas, time of first breakthrough, total process time and width and speed of the reaction zone, showing how these depend on the operating conditions and material parameters. The relations show clearly how to  optimise the carbon capture process. By comparison with experimental data the analytical solution may also be used to calculate unknown system parameters.
\end{abstract}


\section{Introduction}

The issue of excessive amounts of carbon in the atmosphere, which is still being added to at an alarming rate, and the resultant effect on the climate is  well-documented. Consequently mankind must look to a range of solutions including a reduction in current production,  removing existing carbon from the environment and safe storage or reuse (in a sensible way). The process of carbon capture falls within these measures and in this paper we will develop and examine a mathematical model for carbon capture by adsorption. The governing equations developed, although similar to previous works, contain the correct terms and coefficients. Further, we present a novel analytical solution
which predicts how the gas concentration and amount of adsorbate vary along the column as functions of the experimental conditions.

The particular process of interest for this study involves a gas being forced through a column packed with an adsorbent material. This is a standard method and the literature abounds with experimental, numerical and theoretical papers.
Due to the complexity of the gas flow around the porous media mathematical models for the process typically involve cross-sectional averaged  equations for the heat in the gas and the solid as well as an advection-diffusion equation for the gas concentration. The equations are linked through a term describing the adsorption rate. In the advection-diffusion equation this appears as a mass sink, while in the solid heat equation it  appears as a source due to the exothermic reaction. Various forms of this type of system may be found in the review papers \cite{BenMansour,Li,Shaf14}, the modelling of  fixed bed experiments  \cite{Dantas} and the high feed rate experiments of \cite{Reza12}. Isothermal models are investigated in the experimental and numerical studies of \cite{AlJanabi,Shaf15}, the numerical study of \cite{Rhag} and the studies on adsorption equilibrium and breakthrough \cite{Shen,Zhao}. Similar models occur in the literature to describe absorption and liquid (as opposed to gas) flow \cite{Xu13, Patel}.

In many publications the averaged governing equations are immediately written down, referring the reader to the book of Ruthven \cite{Ruthven}. In the following we will start at an earlier point, using the standard heat and advection-diffusion equations and then carrying out the averaging process. Through this method we are able to identify a number of errors and inconsistencies which have propagated through the literature. We discuss these in detail later and also explain why despite the errors they still permit good agreement with experiment.

In the following section we will present the system of equations describing the cross-sectional averaged temperature and concentration in a circular column. In \S \ref{NonDimSec}, by non-dimensionalising the equations,
we are able to identify dominant terms and also the scales describing the general features of the reaction. We then compare the model results with data from the pressure swing adsorption experiments of  \cite{Shaf15}. The final section concerns the derivation of an exact analytical solution, valid away from the inlet. This solution has important consequences concerning our understanding of the physical process. If all system parameters are known this solution permits the prediction of  quantities such as the first breakthrough time, width of the reaction zone and adsorbed mass. If certain parameters are unknown comparison with breakthrough data permits their determination: in this study using only a few data points we calculate the density of adsorbate and the adsorption rate.

\section{Mathematical modelling}

\begin{table}
\centering
\caption{Nomenclature}
\label{tab:tin}
\begin{tabular}{ccc}
\hline
& Symbol & Dimension \\
\hline
Specific heat capacity at constant pressure & $c$ & J/ (kg K)\\
Diffusion coefficient of component $i$ & $D_i$ & m$^2$/s \\
Heat transfer coefficient & $h$ & W/ (m$^2$ K) \\
Concentration of component $i$ & $C_i$ & mol/m$^3$ \\
Thermal conductivity & $\lambda$ & W/(m K)\\
Pressure & $p$ & Pa \\
Amount of adsorbed solid & $q$ & mol/kg \\
Inner column radius & $R$ & m \\
Universal gas constant & $R_g$ & J/ (mol K)\\
Gas temperature &$T$ & K \\
Wall thickness & $t_w$ & m \\
Velocity & $\mathbf{u}= (u(x,r), w(x,r))$ & m/s \\
Gas molar fraction & $y$ & - \\
Adsorption heat & $\Delta H$ & J/mol \\
Bed void fraction & $\epsilon$ & - \\
Dynamic viscosity & $\mu$ & Pa s \\
Composite temperature & $\phi$ & K \\
Solid temperature & $\theta$ & K \\
Density & $\rho$ & kg/m$^3$ \\
\hline
\textbf{Subscripts }& & \\
Ambient & $a$ & -\\
Gas & $g$ & -\\
Solid & $s$ & -\\
Adsorbed & $q$ & -\\
Wall & $w$ & -\\
\hline
\end{tabular}
\end{table}

\begin{figure}
\centering
\includegraphics[width=0.6\textwidth]{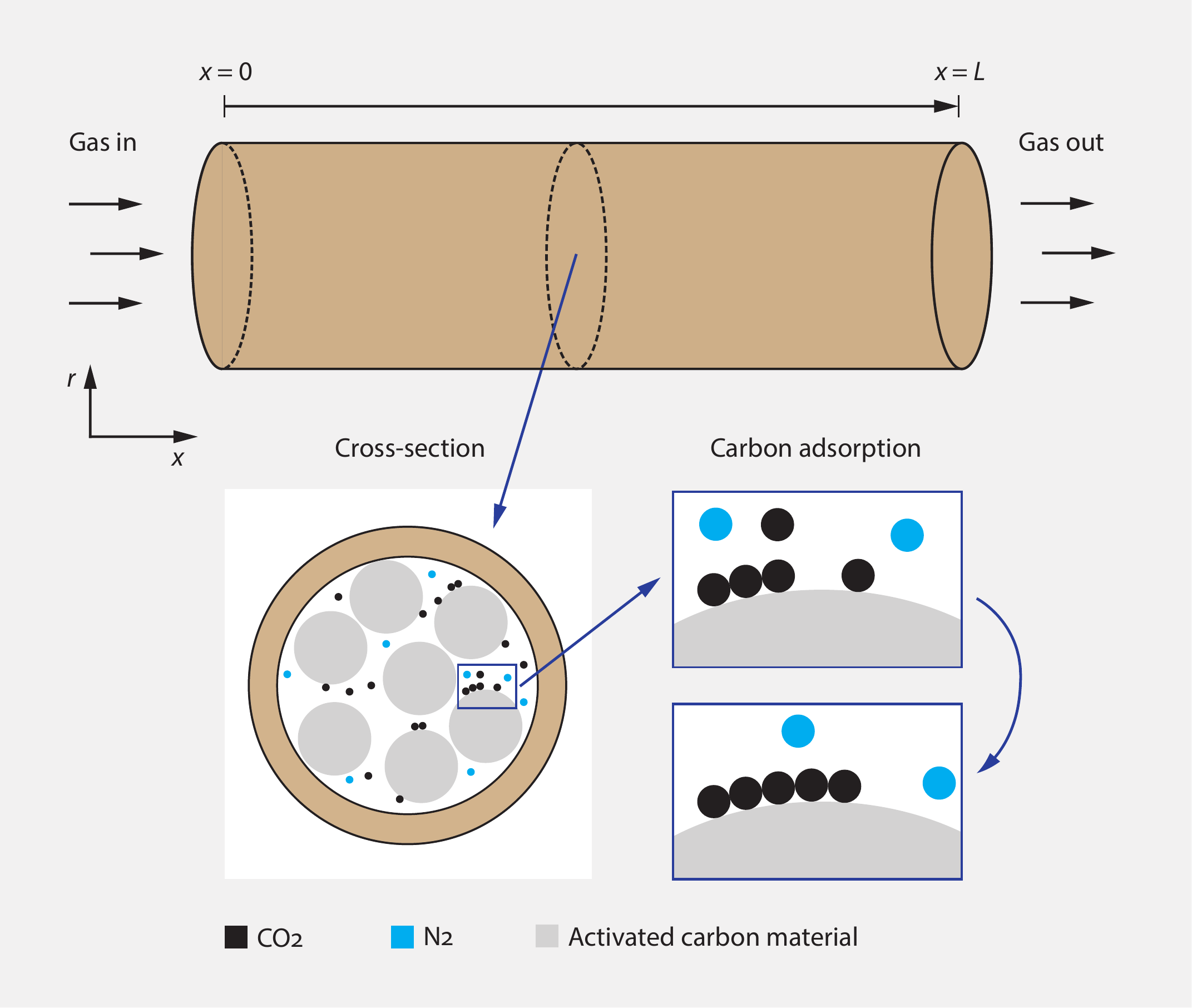}
\caption{ Illustration of the carbon capture by adsorption process. A gas containing CO$_2$ (typically CO$_2$ and N$_2$ in standard experimental configurations) passes through a cylindrical column bed of adsorbent, which adsorbs
	CO$_2$ molecules onto its surface or within its pores.  }
\label{fig:illu}
\end{figure}

The typical experimental set-up involves a circular cross-section column containing an adsorbing material which is placed inside an oven or furnace to regulate the temperature. Gas is passed through the column and the concentration measured at the outlet.  {A schematic of the experimental set-up is presented in Fig.~\ref{fig:illu}.} In the following we will take data and parameter values from \cite{Shaf15}, which involves a CO$_2$, N$_2$ mixture passing through a bed of activated carbon. A full description of the experiment may be found in \cite{Shaf15}. Similar experiments are described in detail in the reviews of \cite{BenMansour,Li,Muker} and the experimental study \cite{Dantas}, for example. The only reason to choose the data of \cite{Shaf15} is that they clearly state all experimental conditions necessary to reproduce their results. The model should work equally well with any other similar experiment.

Here we summarise the assumptions that will be made during the model derivation:
\begin{enumerate}
\item The amount of adsorbate is much less than existing solid, or fills up gaps in the porous media, so the bed void fraction, $\epsilon$, remains constant.

\item Heat released by adsorption is directly transferred to the solid.

\item The adsorption saturation is constant during the process (this will be discussed later).

\item The problem geometry is axisymmetric.

\end{enumerate}

\subsection{Governing equations}

Since the domain under study consists of a packed porous media it is complex and varies along the column. To treat the governing equations  in any sensible way they must be averaged hence it is standard to apply a cross-sectional average. Not all cross-sections will have the same fraction of solid to gas so strictly we are thinking in terms of an ensemble average, which would be a typical average over a number of cross-sections.

The average gas temperature is defined as
\bea
\label{AvT}
\epsilon \pi R^2 \widetilde{T}  =  2 \pi \int_{0}^{R}  T r ~dr ~ , ~ &\Rightarrow& ~  \widetilde{T}  = \frac{2}{ \epsilon  R^2} \int_{0}^{R} T r ~dr~ ,
\eea
where $T$ is the temperature in the gas and $\epsilon$ the void fraction.
In the same manner we average the solid temperature, $\theta$, concentration of gas, $C$, and the amount adsorbed, $q$,
\bea
\label{Avth}
\widetilde{\theta}  = \frac{2}{(1-\epsilon) R^2} \int_{0}^{R} \theta r ~dr ~ ,~~~~~~
\widetilde{C}  = \frac{2}{\epsilon R^2} \int_{0}^{R} C r ~ dr ~ ,&&  ~~~~~~
\widetilde{q}  = \frac{2}{(1-\epsilon) R^2} \int_{0}^{R} q r ~dr ~ .
\eea
Other parameters are defined in the Nomenclature table.
In the $T$ integral the integrand is only non-zero over an area $\epsilon \pi R^2$ of the cross-section. Similarly, the gas occupies the same region. The solid (and hence material adsorbed onto the solid) occupies an area $(1-\epsilon) \pi R^2$.
The omission of the radial angle indicates that we restrict our attention to an axisymmetric cross-sectional average.
A more rigorous way to deal with the averaging through the column would be to follow an averaging process such as that described in \cite{Fowler}. This requires introducing an indicator function $\chi$, where $\chi=1$  corresponds to the gas phase and $\chi=0$ the solid. We could then integrate the governing equations, multiplied by $\chi$, over a given volume, leading to volume averaged quantities. In regions where $\chi$ switches values a law for the interchange between phases must be applied. In the Supplementary Material we apply a specific form of this technique, that is we restrict our attention to an axisymmetric cross-sectional average. The choice of axisymmetry avoids the need for a more complex mathematical analysis and coincides with the equations studied in the carbon capture literature.

The averaging process leads to the following equations describing the heat flow,
\begin{align}
\label{gasheatav}
(\rho c)_g \left(\pad{\widetilde{T}}{t} +  \pad{(u\widetilde{T})}{x} \right) & =
\lambda_g \padd{\widetilde{T}}{x} +
\frac{2 \widetilde{h}_{gs}}{\epsilon  R} \left(\widetilde{\theta}-\widetilde{T}\right)  + \frac{2 h_{wg}}{\epsilon R} \left(T_w-\widetilde{T}\right)
- \frac{p}{\epsilon} \pad u x ~,\\
\label{solidheatav}
(\rho c)_s \pad{\widetilde{\theta}}{t} & = \lambda_s\padd{\widetilde{\theta}}{x} + \frac{2 \widetilde{h}_{gs}}{(1-\epsilon)  R} \left(\widetilde{T}-\widetilde{\theta}\right) + \frac{2 h_{ws}}{(1-\epsilon) R}  \left( T_w-\widetilde{\theta}\right) + \sum_i \Delta H_i \rho_{qi}  \pad{\widetilde{q}_i}{t} \, .
\end{align}
The parameter  $c$ is the \emph{specific heat capacity measured at constant pressure}, this will be discussed later.

The solid heat equation contains a source term due to the heat released as the gas is adsorbed. An important point to note here is that the rate of heat generation is proportional to $\rho_{q} \Delta H$, where $\rho_{q}$ is the density of the adsorbed material (adsorbate), this is typically much lower than that of the adsorbing material (adsorbent) \cite{Baha}.  In fact the whole source term is an approximation, the correct form would link the heat flow to an extra energy balance describing the creation of new solid (a Stefan condition). This may be seen in studies of ablation, see \cite{Mitch08} for example. To avoid solving the full moving boundary problem at each interface, further complicating the solution it is standard to add the heat generated to the solid temperature. Heat exchange with the gas is then accounted for by a convective boundary condition.

When the gas and solid are taken to have different temperatures it is termed the \lq non-equilibrium thermal model' \cite{Li}. If we set $\widetilde{T}=\widetilde{\theta}=\widetilde{\phi}$ then the \lq equilibrium thermal model' is found. A single equation for $\widetilde{\phi}$ may be obtained by adding $\epsilon$ times equation \eqref{gasheatav} to $(1-\epsilon)$ times equation  \eqref{solidheatav}
\begin{align}
(\rho c)_p  \pad{\widetilde{\phi}}{t} + \epsilon (\rho c)_g  \pad{(u \widetilde{\phi})}{x}  =
k_p \padd{\widetilde{\phi}}{x} &+ \frac{2(h_{wg}+h_{ws})}{ R} (T_w-\widetilde{\phi})   + (1-\epsilon) \sum_i \Delta H_i \rho_{qi}  \pad{\widetilde{q}_i}{t} - p \pad u x \, ,
\end{align}
where the subscript $p$ denotes the porous medium ($z_p = \epsilon z_g + (1-\epsilon) z_s)$.

The average concentration, $\widetilde{C}_i(x,t)$, is  described by
\bea
\label{ConcEqav}
\pad{\widetilde{C}_i}{t} +  \pad{(u\widetilde{C}_i)}{x} = D_i \padd{\widetilde{C}_i}{x} -  \frac{1-\epsilon}{\epsilon} \rho_{qi}  \pad{\widetilde{q}_i}{t}   ~ .
\eea
The coefficients $D_i$ represent an 'effective axial dispersion coefficient' which lumps all mechanisms contributing to axial mixing
(e.g. molecular diffusion or turbulent mixing as flow passes round particles and recombines) \cite[P. 208, 209]{Ruthven}. Since it is dominated by the flow if there is more than one component to the gas it may be assumed to have
the same value for each component. Note, this equation only holds where $C_i > 0$, i.e. for $x \in [0, s_i(t)]$ where $x=s_i(t)$ denotes the boundary beyond which the concentration is zero.
Since $u(x)$ represents the velocity at which energy is being advected by the gas it must be interpreted as the interstitial velocity (rather than the superficial velocity). If the total gas flux at the column entrance is $Q_0$ then $u(0) = Q_0/(\epsilon \pi R^2)$.

\subsection{Boundary and initial conditions}\label{BCSec}

At the inlet the gas has concentration $C_{i0}$ of component $i$, just inside the column it moves with velocity $u$ and is adsorbed by the porous media. The velocity just outside the column will be the same as that just inside, and occupying the same area (any gas above a solid section will have zero velocity and so not contribute to the gas flux). Hence we apply a Dankwert condition to the average concentration
\bea
\label{InletCond}
u C_{i0}  = \left. \left(u \widetilde{C}_i  - D_i \pad{\widetilde{C}_i }{x}\right) \right|_{x=0} \, .
\eea
At the outlet $x=L$ a similar balance holds, however it is standard to assume that $\widetilde{C}_i(L^-,t) \approx C_{i}(L^+,t)$ and so
\bea
\label{CLcond}
\left. \pad{\widetilde{C}_i(L^-,t)}{x}\right|_{x=L} = 0 \, .
\eea
However, we note that before breakthrough not all gas components reach the outlet. In this case we impose
\bea
\pad{\widetilde{C}_i(s_i,t)}{x} = \widetilde{C}_i(s_i,t) = 0 \, ,
\eea
where $x=s_i(t)$ is the point where the $i$-th gas concentration reaches zero.
The condition \eqref{CLcond} has been replaced by two conditions. The additional condition is necessary to determine the new unknown $s(t)$.

For the temperature we may carry out a similar balance, in this case for the energy flux
\bea
\label{InletCondT}
u  (\rho c)_g T_a  = \left. \left( u  (\rho c)_g \widetilde{T}  - \lambda_g \pad{\widetilde{T} }{x}\right) \right|_{x=0} \, .
\eea
As above, at the outlet we impose
\bea
\left.  \pad{\widetilde{T} }{x} \right|_{x=0}= 0 \, .
\eea
Similar conditions hold in the solid.
In experiments where the bed has been saturated by N$_2$ the initial condition for N$_2$ is approximated with an ideal gas law
\bea
\widetilde{C}_i(x,0) = \frac{p}{R_g \widetilde{T}}~ , \qquad \widetilde{q}_i(x,0) = \tilde{q}_i^* \, .
\eea
The CO$_2$ initial condition for N$_2$ saturated and unsaturated scenarios should be
\bea
\widetilde{C}_i(x,0) =  \widetilde{q}_i(x,0) = 0 \, .
\eea

\subsection{Other system variables}

There exist a variety of formulae to describe the adsorption rate, but possibly the most common is
a linear kinetic relation,
\bea
\pad {\widetilde{q}_i}{t} = k_i (\tilde{q}_i^* - \widetilde{q}_i) \, ,
\eea
where $\widetilde{q}_i$ is the averaged concentration within the adsorbing particle.
This form is termed a Linear Driving Force model in the literature. Other forms such as the second order, Avrami and pore diffusion relations are discussed in \cite{Li,Shaf15}. The premise of the LDF is that
the rate of change of a species is directly proportional to the difference between its
saturation concentration and the mean
concentration within the particle. The model is very simple and contains only two parameters, $k_i$ is a rate constant and $\tilde{q}_i^*$ is the
maximum possible value for the average concentration. This latter value differs from values measured at a surface due to the averaging process $\tilde{q}_i^* = 2 \tilde{q}^*_{surf}/((1-\epsilon) R^2)$. A number of \emph{isotherm} models exist for its estimations such as Langmuir \cite{Li}, Langmuir-Freundlich \cite{Wood} and Toth \cite{Shaf15}. Since we take experimental data from \cite{Shaf15}, we will use their formula which takes the form
$\tilde{q}_i^* = q_c + q_p$,
which incorporates chemical and physical adsorption mechanisms. Both components have the same form, which is identical to the Toth isotherm,  but with different exponents. Details are provided in  \ref{qSec}.

An important point to note is that throughout the derivation we have written $\partial \widetilde{q}_i/\partial t$ rather than $d \widetilde{q}_i/ dt$ as in all previous works. We use the partial derivative since  $ \widetilde{q}_i$ also depends on position. This becomes clear if we
consider two points within the domain, $x_1, x_2$ where $x_2>x_1$. The column at $x_1$ has had more time to adsorb the gas and so $ \widetilde{q}_i(x_1, t) >  \widetilde{q}_i(x_2,t)$. If we take the specific example with $x_1 = 0$ and $x_2=s$, then at any time $t>0$, $\widetilde{q}_i(0,t) >0$ while $ \widetilde{q}_i(s,t)=0$:  $\widetilde{q}_i$ is therefore dependent on position. The $x$ dependence is also apparent from the numerical results presented later.

The average fluid velocity in a packed bed may be expressed by the Ergun equation
\bea
\label{Ergun}
-\pad p x \approx \frac{\Delta p}{L} = 150 \frac{\mu_g}{d_p^2}\frac{(1-\epsilon)^2}{\epsilon^2} u + 1.75 \frac{\rho_g}{d_p}\frac{(1-\epsilon)}{\epsilon} u^2 \, .
\eea
This relation is often quoted in terms of the superficial velocity $v_s = \epsilon u$ and so has slightly different factors.
The terms on the right hand side represent pressure loss due to viscous and kinetic terms respectively. The Ergun equation can deal with both turbulent and laminar flows. If the second term of the right hand side is neglected then the Carman-Kozeny equation for laminar flow is retrieved.

For pressures up to a few atmospheres an ideal gas law may be employed
\bea
\widetilde{C} = \frac{p}{R_g T}~ ,
\eea
and then $\widetilde{C}_i = y_i \widetilde{C}$ where $\widetilde{C}=\sum_i \widetilde{C}_i$ and $y_i$ is the volume fraction.

\section{Differences with previous models}

The system presented above, consisting of the governing equations (\ref{gasheatav}, \ref{solidheatav}, \ref{ConcEqav}) and the boundary conditions of \S \ref{BCSec},
contains a number of differences to standard models in the literature. Here we focus on a few highly cited papers and reviews to highlight these issues. However, the comments below apply to many other papers.

In the present study we write the mass sink term $\rho_q \partial  \widetilde{q}/\partial t$. The density $\rho_q$ is that of the adsorbate which is generally   significantly less than that of the adsorbent $\rho_q < \rho_s$ \cite{Baha}. The energy release associated with this process  results in a source term proportional to the mass sink, $\Delta H \rho_q \partial  \widetilde{q}/\partial t$. It is common practice to represent the sink by $\rho_s d  \widetilde{q}/d t$. Given that $\rho_s > \rho_q$ employing $\rho_s$ could predict a significantly faster reduction in the gas concentration than in practice and a higher temperature increase. In \cite{BenMansour,Li,Shaf14,Dantas,Shaf15,Shen,Dantas2,Simo}  the solid particle density, $\rho_s$ is used in both the concentration and heat equations. In \cite{Bhatt,Plaza,Reza09} the bulk density is employed. However, in  \cite{Ruthven}, the oft quoted source of the governing equations, the variable considered is in fact the product of  density and adsorbate, so avoiding the issue. In \S \ref{TWsec} we show that in the experiment studied in this paper $\rho_{q} \approx 325$\,kg/m$^3$, which is approximately one sixth of the particle density $\rho_s = 1818$\,kg/m$^3$.

The standard heat equation quoted in studies such as \cite{BenMansour,Li,Shaf14,Dantas,Wood,Dantas2} neglects heat conduction in the solid. In fact the standard derivation of this heat equation stems from  a basic energy balance for a single isolated particle, see \cite{Ruthven}, which is then assumed to hold throughout  the column. Due to the simplicity of  the balance, heat transfer between the solid and wall is neglected. For the gas, heat conduction and  transfer at the wall is always included. In \cite{Harder} values for the thermal conductivity of various forms and porosities of activated carbon are given, typically $\lambda_s$ is of the order 0.4 W/m K. For gases the thermal conductivity is typically an order of magnitude lower,  $\lambda_g \sim 0.02 $W/m K, suggesting that conduction in the solid is more important than in the gas. Without further analysis or justification it then seems inappropriate to neglect the solid conduction while retaining that in the gas. This is discussed in the following section.

Confusion arises since $c$ is often simply referred to as the heat capacity, rather than the value measured at constant pressure or constant volume. For a solid the two values are virtually identical: the difference depends on the thermal expansion. For a compressible gas the difference is large. For pressure swing adsorption experiments the pressure is approximately constant at any given point during the adsorption phase, hence the value at constant pressure should be employed.  In certain works both forms, constant volume and constant pressure, are employed, one multiplying the time derivative the other the advection term, see \cite{Dantas,Dantas2,BenMansour17} for example, this is incorrect.

The pressure-work term proportional to  $\partial u/\partial x$ is neglected in most studies, this is consistent with the assumption that mass loss is small compared to the total flow of mass. In the results section we will also apply this approximation.  However, if this term is included in the model then to be consistent  the velocity must also be retained inside the advection derivative, i.e. in the terms $(uC)_x, (uT)_x$. For example in \cite[Supplementary Material]{Plaza} the gas energy equation is based on constant $u$ for the advection term but varying $u$ for the pressure work.
In many examples around 20\% of the gas is removed, calling the assumption on small mass loss into question. However, it is clearly an understandable first approximation for preliminary studies, provided the equations are written in a consistent manner.

A final issue is that a number of authors write down equations involving radial flow, heat exchange at the boundaries (both at the wall and between the gas and solid) and mass loss from the gas concentration, see \cite[Eqs. (19)-(20), (25)-(27)]{Li}, \cite[Eq. (1)]{Muker}.
The inclusion of the heat exchange terms in the governing equations (as opposed to in the boundary conditions) and the mass sink term in the concentration equation is a direct result of the cross-sectional averaging, which eliminates radial variation: the presence of both radial derivatives and boundary exchange terms  is therefore incompatible and should not be used.

A number of these issues will be discussed in more detail later.

\section{Non-dimensional analysis}\label{NonDimSec}

Non-dimensionalisation often allows a system of equations to be simplified, particularly by identifying dominant and negligible terms.
It can also determine the controlling parameters as well as time and length-scales for the process. To correctly carry out the non-dimensionalisation we must focus on a specific example, with realistic parameter values. Hence from now on    we will   focus on a   two component system  composed of CO$_2$ and N$_2$, following the experimental work of \cite{Shaf15}. Specifically we will use data from the experiments of CO$_2$ adsorption
on material \lq OXO-GAC' at 303.15K, appropriate parameter values are listed  in Table \ref{tab:Table1}.
From now on we will use  $C$ to denote the CO$_2$ concentration. For simplicity we will also assume $u$ to be constant, i.e. the rate of mass removal is small compared to the total mass flow. The equilibrium saturation $q^*$ is calculated from \cite[eq. (7)]{Shaf15}, which accounts for both physical and chemical adsorption mechanisms, details are provided in  \ref{qSec}. The interstitial velocity is $u = Q/(\epsilon\pi R^2)$. The apparent density, $\rho_a = 800$\,kg/m$^3$, quoted in \cite{Shaf15} indicates a particle density $\rho_s \approx \rho_a/(1-\epsilon) \approx 1818$ kg/m$^3$. This density is used in their mass sink term. In our calculations we employ $\rho_{q} = 325$ kg/m$^3$. Instead of their estimate $k_1 = 0.0315$\,s$^{-1}$, we use $k_1 = 0.0137$ s$^{-1}$. These values will be discussed later. Most authors calculate the axial diffusion through a formula involving the molecular diffusivity, Schmidt number and Reynolds number, see \cite{Dantas,Shaf15} for example. In \cite{Dantas} this leads to values ranging between $10^{-5}$ to $10^{-3}$. Here we simply use a value taken from \cite[Table 16.2-2]{BSL} and adjust due to the porous nature of the solid, $D = 1.44 \times 10^{-5}/\epsilon$, this will also be discussed later.

\begin{table}
\centering
\begin{tabular}{cccc}
\hline
& Symbol & Value & Dimension \\
\hline
Initial concentration (CO$_2$)  & $C_{0}$ & 6.03 & mol/m$^3$ \\
Adsorption saturation   & $\tilde{q}^*$ & 1.57 & mol/kg \\
Bed void fraction & $\varepsilon$ & 0.56 & - \\
Density  of adsorbed CO$_2$ & $\rho_{q}$ & 325 & kg/m$^3$ \\
Axial diffusion coefficient & $D$ & $2.57 \times 10^{-5}$ & m$^2$/s \\
Volume fraction (CO$_2$) & $y_1$ & 0.15 & - \\
Bed length & $L$ & 0.2 & m \\
Pressure & $p$ & $101325 (1) $ & Pa (Atm) \\
Flux & $Q$ & $8.3\times 10^{-7}$ & m$^3$/s \\
Bed radius & $R$ & 0.005 & m \\
Temperature (for $\tilde{q}^*$ calculation) & $T$ & 303.15 & K \\
Interstitial velocity  & $u$ & 0.019 & m/s \\
Adsorption rate constant (CO$_2$) & $k_1$ & 0.0137 & s$^{-1}$ \\
Specific heat capacity of solid/gas & $c_s/c_g$ & 880/1000 & J/(kg K)\\
Solid/gas density & $\rho_s/\rho_g$ & 1818/1.2 & kg/m$^3$\\
Thermal conductivity & $\lambda_s/\lambda_g$ & 0.4/0.026 & W/(m K)\\
Heat of adsorption & $\Delta H_1$ & $2.18 \times 10^{4}$ & J/kg \\
\hline
\end{tabular}
\caption{Values of the thermophysical parameters mainly taken from \cite{Shaf15}, except $k_1, \rho_{q}$, solid thermal properties are taken from \cite{Harder,Uddin,Zhequan}. Gas thermal properties are those of air at 300K and 1 atmosphere.}
\label{tab:Table1}
\end{table}

We scale concentration with the initial value of CO$_2$, the amount of adsorbent will then be scaled with $\tilde{q}^*$ (which for the present study is assumed to be constant). Length and time-scales will be left unspecified for the moment
\bea
\widehat{C} = \frac{\widetilde{C}}{C_{0}}~ , \qquad \hat{x} = \frac{x}{\cal{L}}~ , \qquad \widehat{t} = \frac{t}{\Delta t}~ , \qquad \widehat{q} = \frac{\widetilde{q}}{\tilde{q}^*} \, .
\eea
The CO$_2$ concentration is then governed by
\begin{align}
\frac{C_{0}}{\Delta t} \pad{\widehat{C}}{\widehat{t}} + \frac{u C_{0}}{{\cal L}} \pad{\widehat{C}}{\widehat{x}} &= \frac{D C_{0}}{{\cal L}^2}   \padd{\widehat{C}}{\widehat{x}} -  \frac{1-\epsilon}{\epsilon} \frac{\rho_{q} \tilde{q}^*}{\Delta t} \pad{\widehat{q}}{\widehat{t}}~ ,  \\
\pad {\widehat{q}}{\widehat{t}} &= k_1 \Delta t\left(1 - \widehat{q}\right)
~ .
\end{align}
Much information may be gained from these equations. For example, the second equation indicates that CO$_2$ adsorption occurs on a time-scale $\Delta t = 1/k_1 = 1/0.0137 \approx 73$\,s. So significant changes occur over time-scales of approximately 1 minute. Since $\rho_{q} \tilde{q}^* \gg C_{0}$ it is clear that
the time derivative of the concentration is negligible compared to the sink term and therefore may, in general, be neglected. Except for when the interstitial velocity is extremely low, $u \sim D/{\cal L}$, advection  dominates over diffusion hence we expect the advection term to balance with the sink term, this necessitates the choice of length-scale
\bea
\frac{u C_{0}}{{\cal L}} = \frac{1-\epsilon}{\epsilon} \frac{\rho_{q} \tilde{q}^*}{\Delta t} \quad \Rightarrow \quad {\cal L} = \frac{\epsilon u C_{0} }{(1-\epsilon)k_1 \rho_{q} \tilde{q}^*} \, .
\eea
Taking the values of Table \ref{tab:Table1}   gives ${\cal L} \approx 2$\,cm: this is the order of magnitude of width over which we expect the reaction to occur, i.e. the region over which the concentration falls from its inlet value to approximately zero (in \cite{Zhao} reaction zones between 0.5 and 3\,cm are quoted for CO$_2$ adsorption on silica gels). We can now see that diffusion becomes important for an interstitial velocity of the order $D/{\cal L} \sim 1.3$\,mm/s. This is in keeping with the unqualified statement in \cite{Zhao} that axial diffusion may be neglected if the flow is not too slow. Here we quantify the term \lq not too slow', i.e. $u \gg D/{\cal L}$.

Using the length-scale ${\cal L}$ and time-scale $\Delta t = 1/k_1$ we obtain
\begin{align}
\label{C1ND}
\delta_1 \pad{\widehat{C}}{\widehat{t}}+ \pad{\widehat{C}}{\widehat{x}} &= Pe^{-1}   \padd{\widehat{C}}{\widehat{x}} - \pad{\widehat{q}}{\widehat{t}} ~ , \\
\label{qthateq} \pad {\widehat{q}}{\widehat{t}} &= ( 1 - \widehat{q})
~ ,
\end{align}
where $Pe^{-1} = D/({\cal L} u)$ is the inverse Peclet number and $\delta_1 = {\cal L}/u\Delta t = {k_1 \cal L}/u$. Here we find $Pe^{-1} \approx 0.066$ and $\delta_1 = 0.015$.
The small size of $Pe^{-1}$ indicates that, in this case, diffusion plays a minor role in the process (as discussed earlier). Consequently a detailed calculation of the value of $D$ is unnecessary. Neglecting this term would lead to errors of the order 7\%, however, retaining it may be useful for a numerical scheme if sharp concentration gradients occur such as at small times. The term $\delta_1$ is even smaller, its neglect will lead to errors of the order 1\%.

Under the standard assumption $\widehat{q}= \widehat{q}(t)$ the integration of \eqref{qthateq} is simple and leads to $\widehat{q} = 1-e^{-\widehat{t}}$, as quoted in \cite{Shaf15,Sarker}. This satisfies the initial condition $\widehat{q}(\widehat{t}=0) = 0$. However, as discussed earlier there is an obvious $\widehat{x}$ dependence in the adsorption term. The issue is easily resolved. If we denote the time for the gas to reach a given point $\widehat{x}$ as $\widehat{t}=\widehat{t}_s(\widehat{x})$ then the correct initial condition at that point is $\widehat{q}(\widehat{x},\widehat{t}_{s}(\widehat{x})) = 0$. Consequently the correct solution to the adsorption equation is
\bea
\label{qanal}
\widehat{q}(\widehat{x},\widehat{t}) = 1- e^{-(\widehat{t}-\widehat{t}_{s})} \, ,
\eea
which holds over the domain where $\widehat{C} > 0$, i.e. $\widehat{x} \le  \widehat{s}(\widehat{t})$.
The time $\widehat{t}_{s}$ could for example be found by matching to experiment (or the numerical solution), i.e. if the time of first breakthrough, $\widehat{t}_{fb}$, is recorded then $\widehat{t}_{s}=\widehat{t}_{fb}$ at $\widehat{x}=\widehat{L}$. The evaluation of  $\widehat{t}_{s}$ is discussed in  \S \ref{TWsec}.

Before breakthrough the CO$_2$ concentration equation holds over $\widehat{x} \in [0,\widehat{s}(\widehat{t})]$ and is to be solved subject to
\bea
\label{CBC1}
1 = \left. \left(\widehat{C} - Pe^{-1} \pad{\widehat{C}}{\widehat{x}} \right)\right|_{\widehat{x}=0}~ , \\
\left. \pad{\widehat{C}}{\widehat{x}}\right|_{\widehat{x}=\widehat{s}} =
\widehat{C}(\widehat{s}, \widehat{t}) = 0 \, .\label{CBC2}
\eea

To investigate the temperature we must define pressure and temperature scales
\bea
\widehat{T} = \frac{\widetilde{T}-T_a}{\Delta T} \, , \qquad \widehat{\theta} = \frac{\widetilde{\theta}-T_a}{\Delta T} \, , \qquad \widehat{p} = \frac{p-p_a}{\Delta p} \, ,
\eea
where $T_a, p_a$ are the ambient temperature and pressure. The scale $\Delta T$ is, as yet, unspecified while $\Delta p$ represents the pressure drop along the column.
Since the rise in temperature is clearly  due to the source term in the solid heat equation we choose $ \Delta T = \Delta H \rho_{q} \tilde{q}^*/((\rho c)_s)$ $\approx 7$\,K. This is consistent with the numerical results
presented  in  \cite[Fig. 9]{Li}, \cite[Fig. 6]{Dantas} where the rise is around 6\,K. Note $\Delta T/T \ll 1$ is an indication that  temperature variation does not play an important role in the process, which then motivates our study of the isothermal problem in \S \ref{ResSec}.
With this choice of $\Delta T$ the solid temperature is  described by
\bea
\label{solidheatNDfIN}
\pad{\widehat{\theta}}{\widehat{t}} = \delta_2 \padd{\widehat{\theta}}{\widehat{x}} + F_1 \widetilde{h}_{gs} \left(\widehat{T}-\widehat{\theta}\right)+ F_1 h_{ws}\left( \widehat{T}_w-\widehat{\theta}\right) + \pad{\widehat{q}}{\widehat{t}} \, ,
\eea
where
\begin{align}
\delta_2 = \frac{\lambda_s \Delta t}{{\cal L}^2(\rho c)_s } \approx 0.04 \, , \qquad  F_1 = \frac{2 \Delta t }{(1-\epsilon)  R(\rho c)_s } \approx 0.04 \,  . \nonumber
\end{align}
Applying the same process to the gas heat equation gives
\bea
\label{gasheatND}
\delta_1 \pad{\widehat{T}}{\widehat{t}} +  \pad{\widehat{T}}{\widehat{x}}   =
Pe_g^{-1} \padd{\widehat{T}}{\widehat{x}} +
F_2 \widetilde{h}_{gs} \left(\widehat{\theta}-\widehat{T}\right)  + F_2 h_{wg}\left(\widehat{T}_w-\widehat{T}\right) ~ ,
\eea
where
\bea
Pe_g^{-1} = \frac{\lambda_g}{ u{\cal L} (\rho c)_g} \approx 0.05 \, , \quad F_2 = \frac{2{\cal L}}{ \epsilon R u (\rho c)_g}  \approx 0.64 \, .
\eea
For further details see the Supplementary Information.

From the non-dimensional form of the heat equations we may obtain a good understanding of the physical process. The solid heat equation shows that the solid temperature increases in time due to the heat generated by adsorption, this is offset by heat loss due to transfer to the walls and gas. The heat transfer coefficient between a slow moving gas and solid is of the order 10 W/m$^2$\,K, hence $F_1 \widetilde{h}_{gs} \approx F_1 {h}_{ws} \approx 0.3$ indicates that around 30\% of the heat may be transferred to the gas or wall. Through the small value of $\delta_2 \approx 0.04$ we see that conduction through the solid has a minor effect. Since $\delta_1 \ll 1$ the gas temperature is almost always in a pseudo-steady state, that is, variations with time are slow compared to the adsorption time-scale. The gas diffusion coefficient $Pe_g^{-1} \approx  \delta_2$, hence diffusion in the solid is just as important as in the gas, so indicating that its neglect  in standard models is inconsistent: \emph{if diffusion is neglected in the solid, for consistency, it should also be neglected in the gas}.
The coefficient $F_2$ is an order of magnitude larger than the solid counterpart $F_1$ which indicates that the gas temperature
primarily varies due to advection and exchange with the solid and walls. The large difference between $F_1$ and $F_2$ shows that the solid plays an important role in giving heat to the gas, while the gas has a smaller effect on the solid temperature, this is a consequence of the  fact that the volumetric heat capacity of the gas is significantly lower than that of the solid, $(\rho c)_g \ll (\rho c)_s$.

\section{Numerical solution}

For the present study we will investigate CO$_2$ adsorption under an isothermal assumption. Consequently our numerical scheme only requires equations to describe the CO$_2$ concentration and the amount adsorbed. The validity of this approximation will be discussed in the results section.

At time $\widehat{t}=0$ the concentration of gas to be adsorbed is zero everywhere inside the column, $\widehat{C}(\widehat{x},0)=0$. At the boundary $\widehat{x}=0$, according to equation \eqref{CBC1}, $\widehat{C}(0,\widehat{t})$ is close to unity (since $Pe^{-1} \ll 1$). This results in a jump in concentration at the start of the process. In practice it is impossible to achieve such a sharp interface but the theoretically imposed jump will clearly cause numerical issues. Consequently in the first part of this section we provide an analytical solution, valid at small times, with which to start the computations.

\subsection{Small time solution}\label{SmallSec}

In the Supplementary Information we demonstrate how, by changing to a short time-scale we are able to determine approximate solutions for both the concentration and amount of adsorbate:
\begin{align}
\widehat{q} &= \widehat{t}-\widehat{t}_e  \, ,\qquad \widehat{s}=1~ , \label{qsmall}\\
\widehat{{C}} & = 1-Pe^{-1} +  Pe^{-1}  e^{-Pe(1-\widehat{x}) } - \widehat{x} \, .\label{Csmall}
\end{align}
These hold for small times such that  $\widehat{t} \ge \widehat{t}_e$, where $\delta_1 \ll \widehat{t}_e \ll 1$. We make the standard choice $\widehat{t}_e=\sqrt{\delta_1}$ which lies in the required range (in the present study this correspond to times of the order 9s). Interestingly, we see that the amount of adsorbate is approximately independent of $x$ at small times. This independence is lost as time progresses.
In practice we start the numerical scheme at time $\widehat{t}=\widehat{t}_e=\sqrt{\delta_1}$, with the profiles defined by (\ref{qsmall}, \ref{Csmall}).

The conclusion that to leading order, for sufficiently small time $\widehat{s}=1$ is a constant may appear strange (and that $\widehat{{C}}$ is independent of time). It represents the distance that the incoming gas travels before being completely used up. Since for small times the adsorbent is relatively fresh the gas can only travel a fixed distance before being completely adsorbed. This will continue until the amount of adsorbate becomes significant and so the adsorbent is less efficient at removing gas. The achieved non-dimensional value $\widehat{s} = 1$  indicates that the length-scale is well-chosen and that it can also represent the distant over which the concentrate travels over fresh adsorbent before being completely used up. An even shorter time-scale is possible, where the gas first enters the column and has not yet reached $\widehat{s}=1$. For numerical purposes (\ref{qsmall}, \ref{Csmall}) are sufficient.

\subsection{Numerical scheme for $t > t_e$}

For most  of the process the problem is a free boundary problem and adsorption only occurs in the growing region $\widehat{x}\in[0,\widehat{s}(\widehat{t})]$ or, in other words, in the region where $\widehat{C}(\widehat{x},\widehat{t})$ is strictly larger than 0. To overcome the numerical difficulty of solving our equations in a growing domain, we rewrite the problem in the form
\begin{align}
\delta_1 \pad{\widehat{C}}{\widehat{t}}+ \pad{\widehat{C}}{\widehat{x}} &= Pe^{-1} \padd{\widehat{C}}{\widehat{x}} - \pad{\widehat{q}}{\widehat{t}}\,,  \\
\pad {\widehat{q}}{\widehat{t}} &= ( 1 - \widehat{q})\, H(\widehat{C})\,,
\end{align}
where $H(\widehat{C})$ represents the Heaviside function
\begin{align}
H(\widehat{C}) =
\begin{cases}
1 \qquad \text{for} \quad  \widehat{C}> 0\,,\\
0 \qquad \text{otherwise}\,.
\end{cases}
\end{align}
The Heaviside function ensures that $\widehat{q}(\widehat{x},\widehat{t})$ only increases in the regions where CO$_2$ is present.
Note, whilst we pointed out earlier that diffusion (also conduction) is small we retain it here firstly to allow us to verify this statement and secondly since it can help smooth the rapid change in concentration due to the initial condition. Although the use of a small time solution removes the actual jump discontinuity.

Because we have effectively removed the moving boundary $\widehat{x}=s(\widehat{t})$, we can modify the boundary condition \eqref{CBC2} to read
\begin{align}
\left. \pad{\widehat{C}}{\widehat{x}}\right|_{\widehat{x}=\widehat{L}} = 0 \, .
\end{align}
We then use standard {second-order} central finite differences in space and explicit Euler in time. {We also employ one-sided second-order finite differences to discretise the derivatives in the boundary conditions, thereby ensuring that the solution is overall second-order in space. The numerical scheme is implemented in Matlab. We choose a time step $\Delta \widehat{t}$ and a mesh size $\Delta \widehat{x}$ that allow the stability condition $Pe^{-1}\Delta \widehat{t}/\delta_1 \Delta \widehat{x}^2<0.5$ to be satisfied and run standard tests reducing $\Delta \widehat{t}$ and $\Delta \widehat{x}$ to ensure that the solution converges. The simulations shown in the present study correspond to $\Delta \widehat{x} = 0.02$ and $\Delta \widehat{t} = 2\cdot10^{-5}$. }



\section{Results}\label{ResSec}

For ease of interpretation we present all results in dimensional form. The concentration is plotted as $\widetilde{C}/\widetilde{C}_0$ where $\widetilde{C}_0= 6.03$ mol/m$^3$, the adsorbate is presented as $\widetilde{q}/\tilde{q}^*$ with $\tilde{q}^* = 1.57$ mol/kg.

\begin{figure}
\centering
\includegraphics[width=0.5\textwidth]{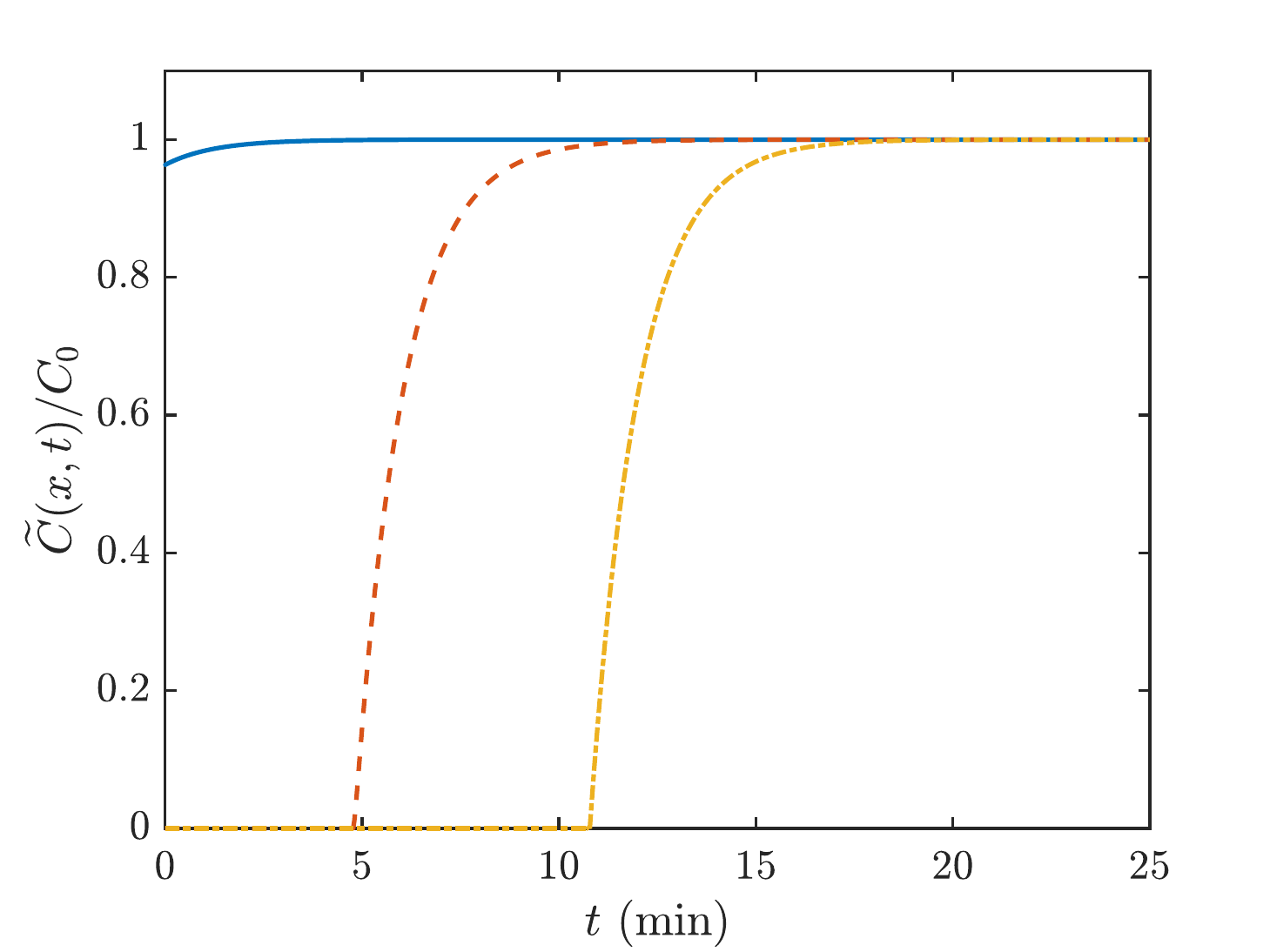}
\caption{Variation of CO$_2$ concentration with time at the inlet (solid), middle (dash) and outlet (dash-dot) of the column. }
\label{fig:Ct}
\end{figure}
In Figure \ref{fig:Ct} we present the variation of CO$_2$ concentration with time at the inlet, middle of the column and outlet. Parameter values are all given in  Table \ref{tab:Table1}. Referring to the solid line we see that at the inlet most of the CO$_2$ passes straight through and after the first few minutes  the gas passes through unchanged. The dashed line shows that CO$_2$ reaches the centre of the column after approximately $ t=4.83$   minutes, at $t=7.73$ minutes it is at 90\% of the inlet value. The same pattern is repeated at the end of the column,the dash-dot curve, the CO$_2$ reaches the end after $t=10.77$ minutes and is at 90\% of the inlet value at 13.57 minutes.

\begin{figure}
\centering
\includegraphics[width=0.5\textwidth]{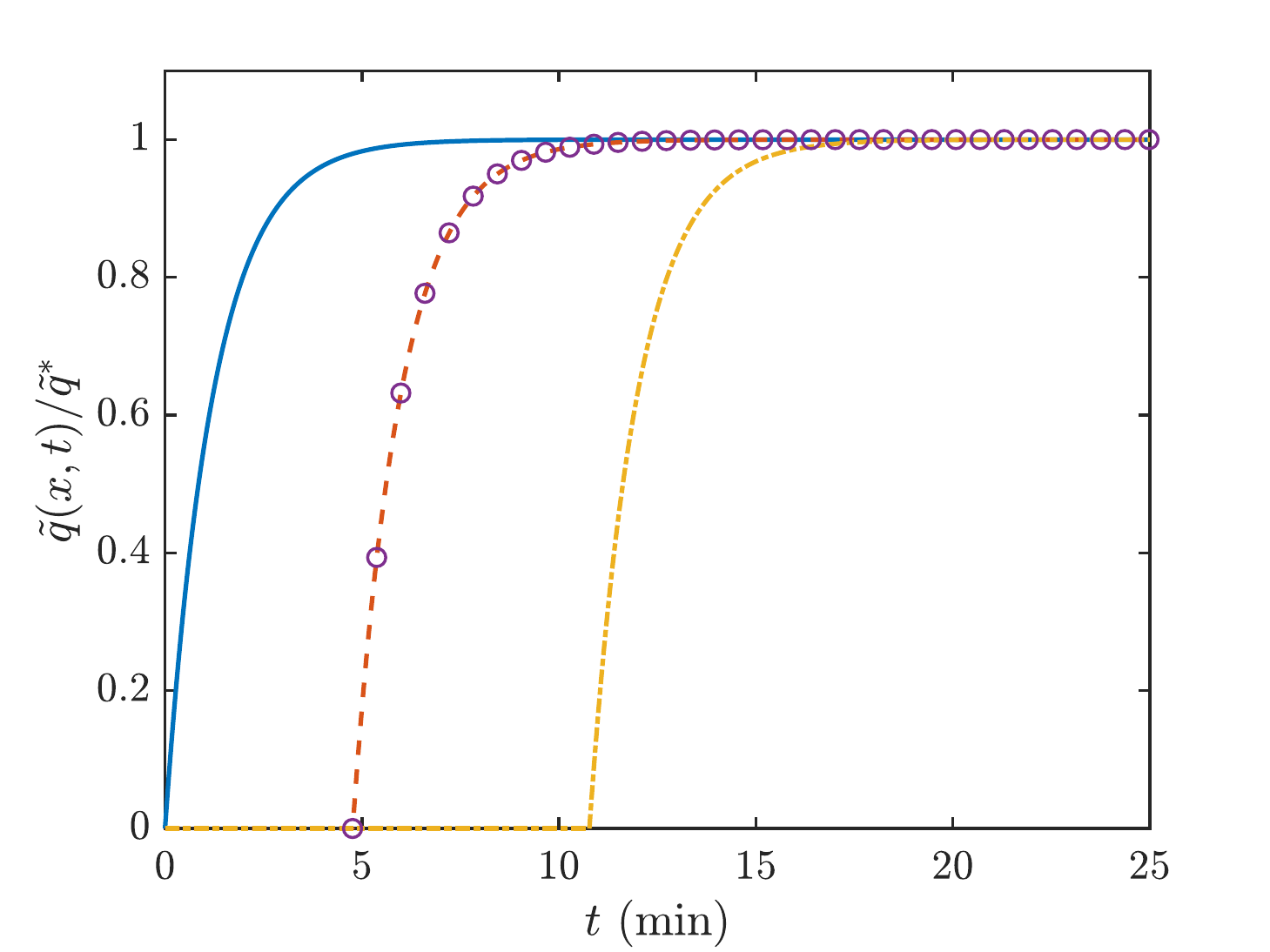}
\caption{Variation of $\widetilde{q}$ with time at the inlet (solid line), middle (dashed line) and outlet (dash-dotted line) of the column. The circles represent the analytical expression for $\widetilde{q}$ of equation \eqref{qadim}. }
\label{fig:Qt}
\end{figure}
Figure \ref{fig:Qt} shows the variation of $\widetilde{q}$ at the inlet (solid), middle (dashed) of the column and outlet (dash-dot). At the inlet $\widetilde{q}$ starts at zero and after 2.95 minutes reaches 90\% of the equilibrium value. This is repeated at the centre and exit of the column, gas first reaches the centre at $4.89$ minutes and the end at $10.92$ minutes and is at 90\% at 7.72 and 13.72 minutes respectively. The behaviour at the centre and exit is very similar to that exhibited by the concentration.
In each case
the time between first reaching a point and being at 90\% of the original value is approximately 2.95 minutes, suggesting that the front moves with a constant speed.
The circles plotted on the figure, show the analytical solution (see eq. \eqref{qanal})
\bea
\label{qadim}
\widetilde{q} = \tilde{q}^* (1-e^{-k_1 (t-t_{s})})\, .
\eea
Here we have chosen
the value $t_{s}=4.89 \times 60$ seconds, so that we may compare the analytical and numerical solutions at the centre of the column. Obviously this is virtually indistinguishable from the dashed line, thus verifying the numerical solution.

\begin{figure}
\centering
\includegraphics[width=0.5\textwidth]{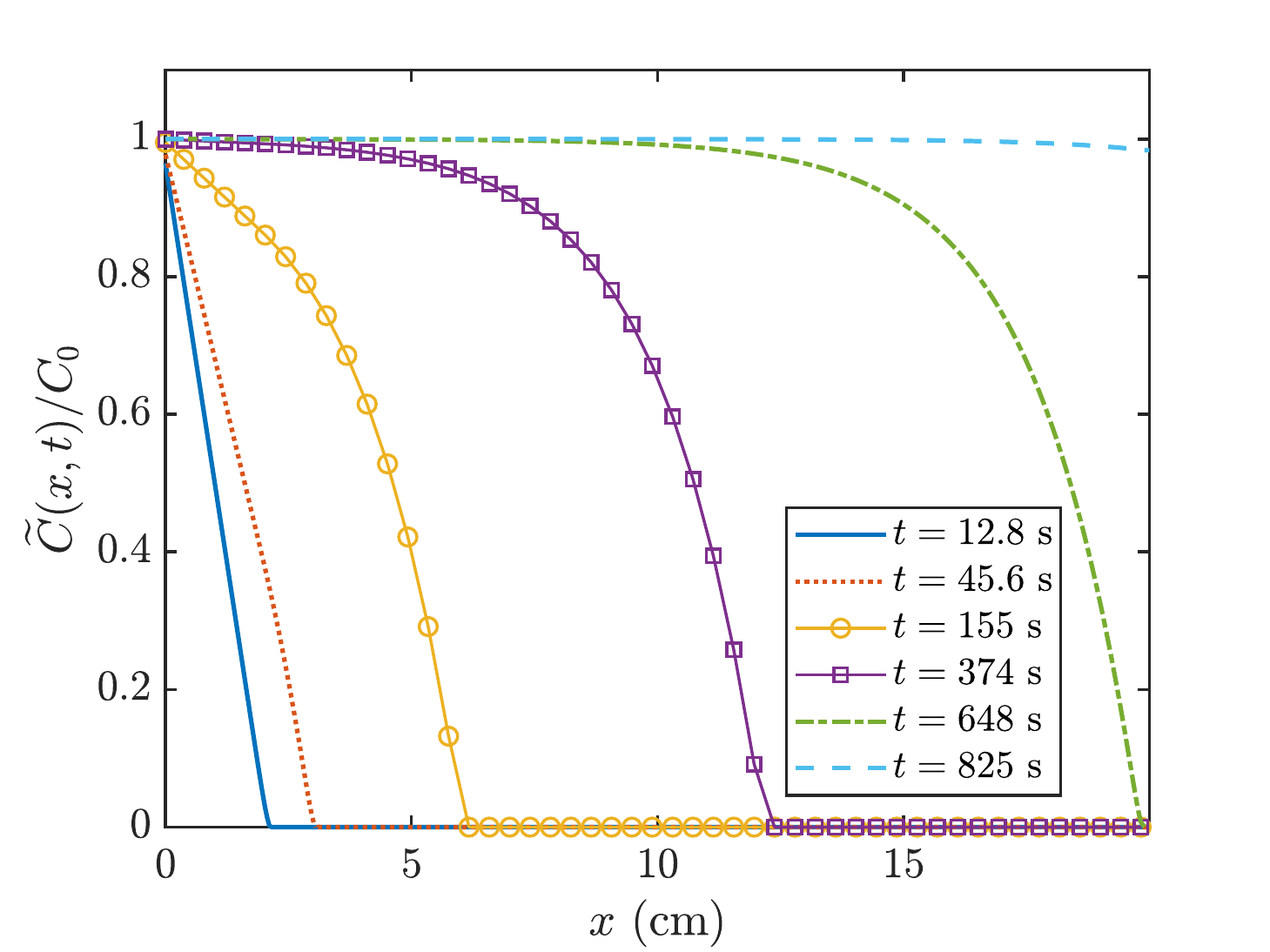}
\caption{The variation of concentration with $x$ at times $t= 12.8, 45.6, 155, 374, 648, 825$\,s.}
\label{fig:6C}
\end{figure}
Figure \ref{fig:6C} shows the concentration along the channel at times $t=$ 12.8, 45.6, 155, 374, 648, 825 s. At very early times, $t=12.8, 45.6$\,s the profile decreases approximately linearly  (more accurately it is close to the form predicted by equation  \eqref{Csmall}). Once the initial transient is over, at  times $t=374, 648$\,s, we see the concentration wave takes on a self-similar form (the drop from 90\% of $C_0$ to zero occurs over a length 4.7\,cm on both curves). This suggests that beyond the initial transient the concentration has a  travelling wave form.
The final time shown, $t=825$\,s is when the concentration at the outlet is 90\% of $C_0$, everywhere else it is above this value and we may consider the process to be effectively complete.

\begin{figure}
\centering
\includegraphics[width=0.5\textwidth]{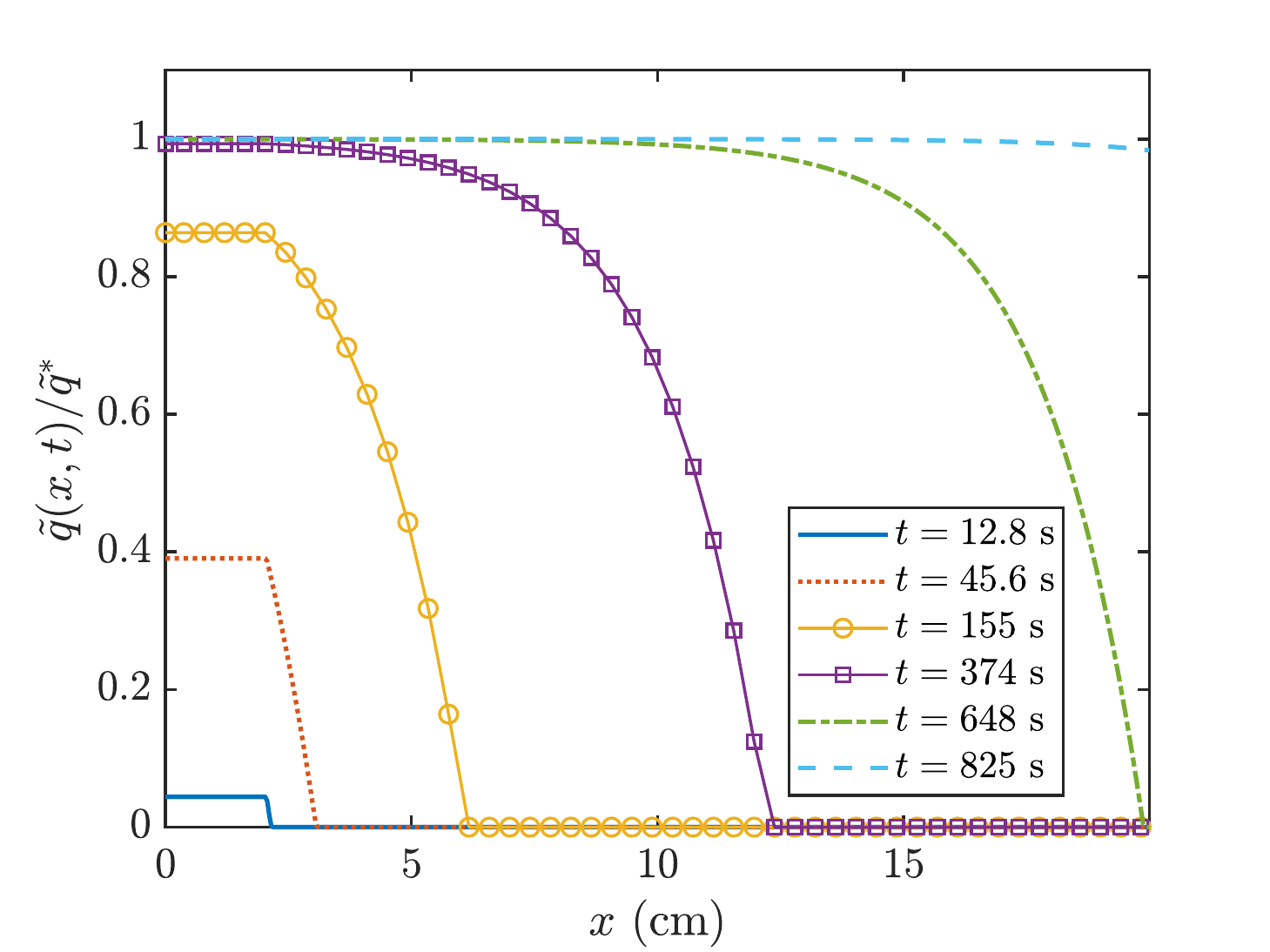}
\caption{The variation of $\widetilde{q}$ with $x$ at times $t= 12.8, 45.6, 155, 374, 648, 825$\,s.}
\label{fig:6Q}
\end{figure}
Figure \ref{fig:6Q} shows the adsorption curves corresponding to the concentration of the previous figure. In \S \ref{SmallSec} we noted that at small times $\widetilde{q}$ is independent of $x$ and grows with time, $\widetilde{q}/\tilde{q}^* \approx k_1( t-t_{s})$. This is quite clear from the results for times $t=$12.8, 45.6s: the curve is approximately flat, followed by a sharp drop to zero.
Once $\widetilde{q}/\tilde{q}^*\approx 1$ at the inlet the $\widetilde{q}$ wave then starts to move in a self-similar form (dropping from 90\% of $\tilde{q}^*_1$ to zero over the same region as the concentration).

A feature apparent from Figures \ref{fig:6C}, \ref{fig:6Q} is that beyond the initial transient the concentration and adsorption curves are hard to distinguish. We will explain this in the following section.

\begin{figure}
\centering
\includegraphics[width=0.5\textwidth]{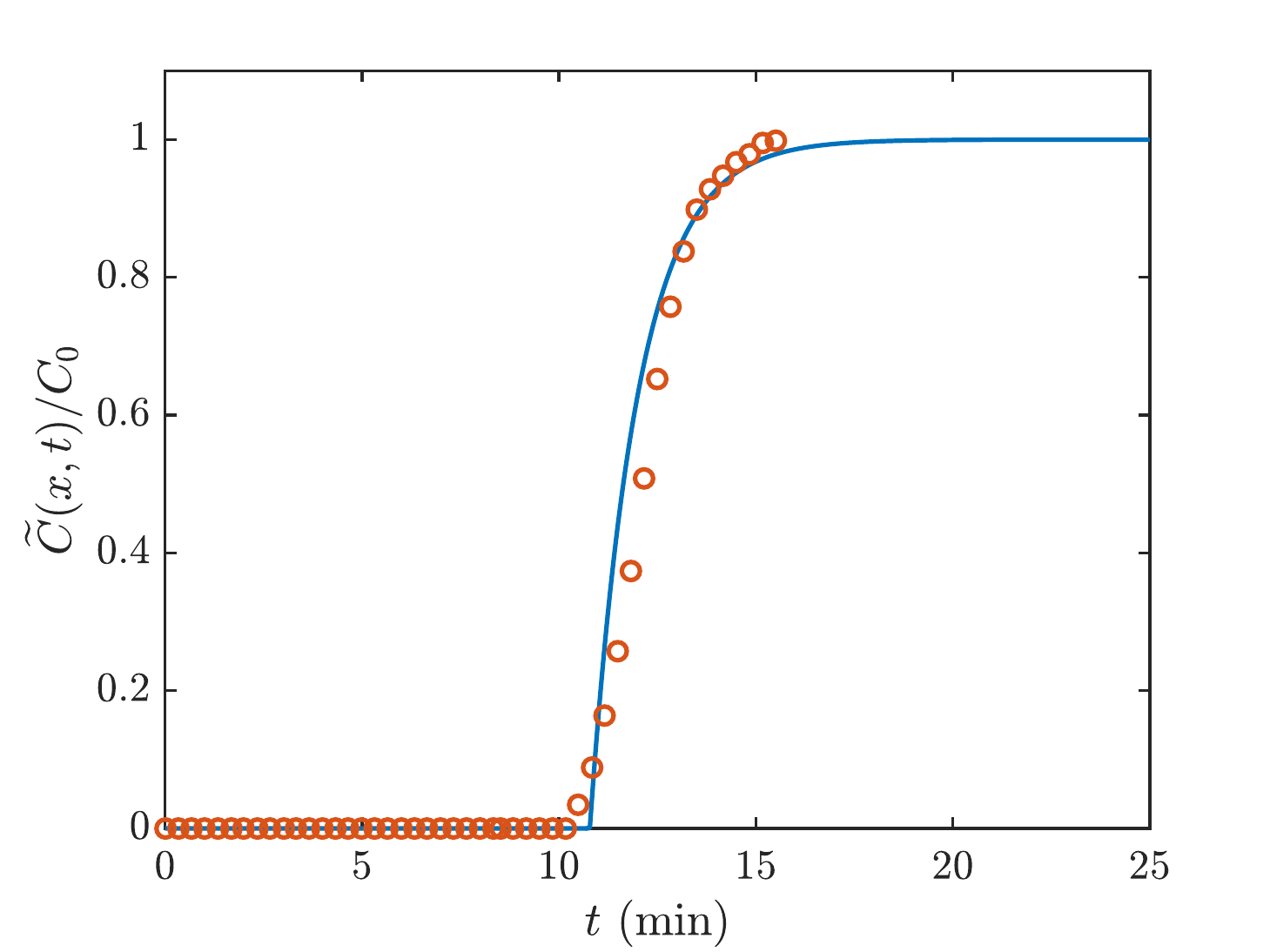}
\caption{Comparison of CO$_2$ concentration at breakthrough from the present isothermal model prediction (solid line) and experimental values taken from  \cite{Shaf15} (circles), for the adsorption of CO$_2$ on activated carbon.}
\label{fig:BT}
\end{figure}
A quantity shown in virtually all experimental papers is the gas concentration at the outlet, the breakthrough curve, which corresponds to the final curve of Figure \ref{fig:Ct}. In Figure \ref{fig:BT} we compare the numerical prediction  with experimental data taken from \cite{Shaf15}. In general the agreement is excellent. The experimental data indicates a slightly earlier breakthrough with a brief slow rise in CO$_2$ concentration, followed by a sharper rise.  This discrepancy may be attributed to the kinetic relation for $\tilde{q}$. Any model where $\widetilde{q}_t \propto \tilde{q}^* - \widetilde{q}$ (where $ \tilde{q}^*$ is constant) will have the highest gradient in the adsorption curve where the gas first meets fresh adsorbent, that is, where  $\widetilde{q}=0$. As observed from the numerical results beyond the initial transient the concentration behaviour is almost identical to the adsorbate, so  the highest concentration gradient will also be at the moving front.

In \S \ref{NonDimSec} we mentioned the usual method of estimating $D$ and how instead of using this we favoured a simple estimate from Bird \et \cite{BSL}. The non-dimensionalisation shows that gas diffusion is described by the inverse Peclet number, which we determined to be small. To verify its small contribution we carried out simulations changing $D$  by factors 0.1 and  10. This had an effect on the concentration near the inlet (since $Pe^{-1} \propto D$ also enters the boundary condition there), however it diminishes rapidly as the gas moves down the pipe. For the case of $0.1 D$ there was no noticeable change to the breakthrough curve, with 10$D$ the time for first breakthrough increased by less than 1\% but the form remained the same. This backs up our assertion that it is not necessary to carry out detailed calculations to determine $D$.

\begin{figure}
\centering
\includegraphics[width=0.5\textwidth]{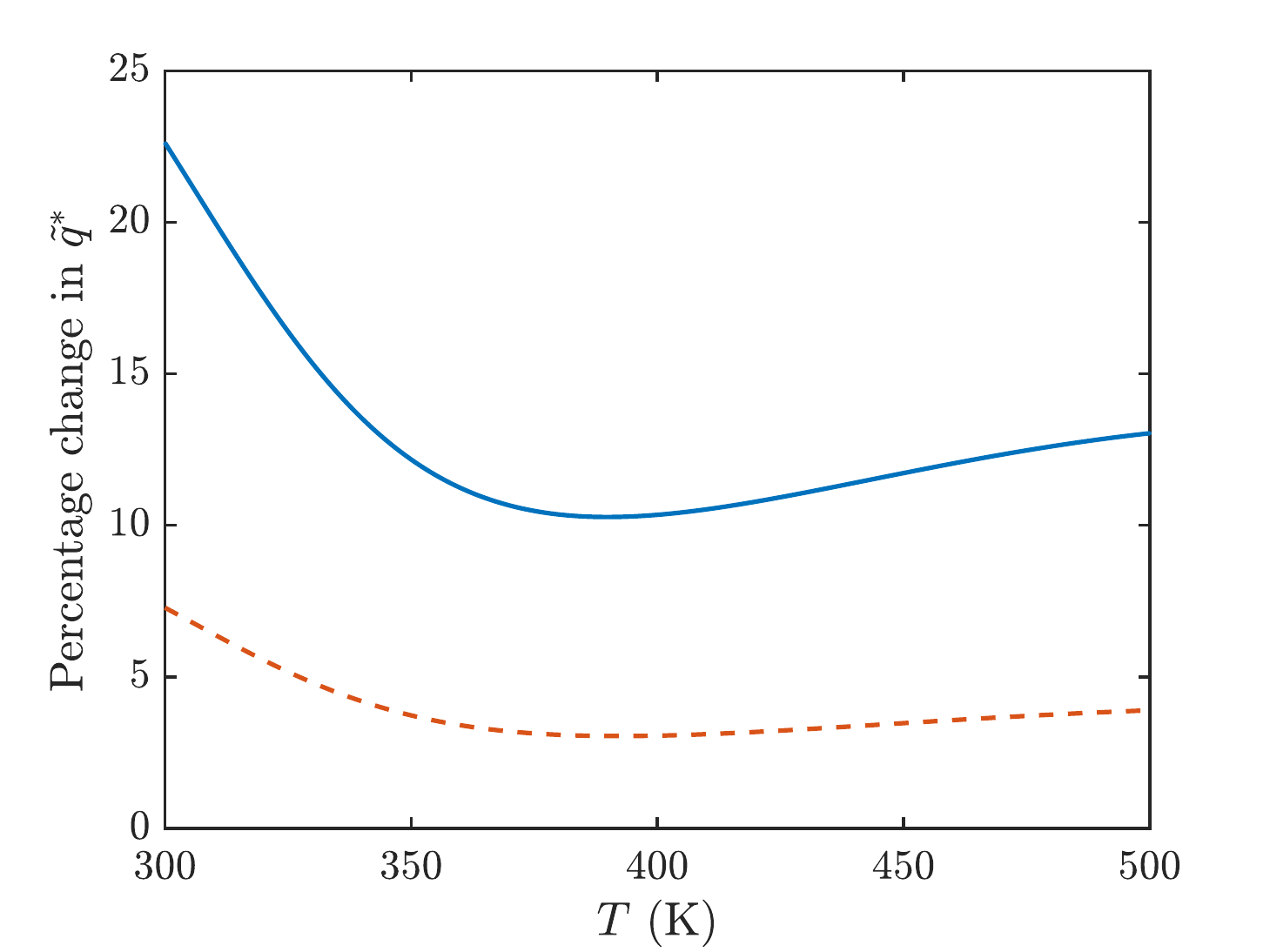}
\caption{Percentage variation in the equilibrium saturation for a temperature rise of 7\,K and 2\,K (solid and dashed lines, respectively).
}
\label{fig:qvar}
\end{figure}

In the non-dimensionalisation process we demonstrated a temperature rise of the order 7\,K. This will have a small effect on the material properties, however, the equilibrium adsorption has a strong temperature dependence, see  \ref{qSec}. For this reason we now examine the effect of temperature on the value of $\tilde{q}^*$. In Figure \ref{fig:qvar}  we show the percentage change $100(\tilde{q}^*(T)-\tilde{q}^*(T+7))/\tilde{q}^*(T)$ and $100(\tilde{q}^*(T)-\tilde{q}^*(T+2))/\tilde{q}^*(T)$. The solid line corresponds to the higher temperature rise and has a maximum just below 23\%. The maximum of the dashed line, which corresponds to the 2\,K rise, is around 7\%.
For the current problem we work at 303\,K, which will result in an approximately 22\% maximum difference in $\tilde{q}^*$ when compared to an isothermal model.
Since the heat is quickly dissipated we would expect this to decrease rapidly. A second source of error is the mass loss, here of the order 15\%, which implies a reduction in velocity ahead of the front of the order 15\%. Since this is outside the reaction zone the velocity there does not affect the calculations. Behind the front the mass loss will vary between 0 and 15\% over a length-scale of around 2cm. An average of 7.5\% mass loss occurring over 10\% of the column suggests an order of magnitude for the error incurred of around 0.75\%. Similarly the temperature variation should lead to small errors, consequently we do not expect either effect to be significant in the analysis but they are certainly worth investigating in future studies.

\section{Travelling wave solution for the gas concentration}\label{TWsec}

The numerical results show that once the initial transient is passed the concentration and adsorption move with a self-similar form, this observation coupled with the constant speed of the front suggests that the reaction may be described by a travelling wave. In this section we will derive
analytical expressions for the shape of the concentration and adsorption curves. These analytical solutions are extremely important to our physical understanding of the process. They provide an accurate description of the wave forms, showing clearly the factors affecting the reaction. At the outlet the solution characterises the breakthrough curve. Comparison with experimental results then permits us to estimate quantities such as the adsorption rate constant $k_1$, the speed and width of the reaction zone and density of adsorbate.

If we consider only the CO$_2$ equation and define a new variable $ \widehat{\eta} = \widehat{x}- \widehat{s}(\widehat{t})$, which has its origin at the reaction front, then the analysis provided in the Supplementary Material leads to the following expressions for
the concentration and adsorption
\begin{align}
\label{Cdim}
\widetilde{C}(x,t)
& \approx  C_0\left(1 -  \alpha e^{\left((x-s_0)/{\widehat{v}\cal L}-k_1 t \right)}\right)  ~ ,\\
\tilde{q}(x,t) & = \tilde{q}^* \left( 1 - e^{((x-s_0)/\hat{v}{\cal L}-k_1 t)}\right) \, ,
\end{align}
where $\widehat{v}=1/(1+\delta_1)$ is constant and
\bea
\alpha =  \frac{\widehat{v}^2 Pe}{\widehat{v}^2 Pe  -1} = \left(1 + \frac{1}{\widehat{v}^2 Pe} +  \frac{1}{(\widehat{v}^2 Pe)^2} \cdots \right) \, .
\eea
The first part of the equality is the exact expression for $\alpha$, the second shows how it may be approximated given that ${\hat v}^2  Pe \approx 15  \gg 1$.
If all terms involving $\widehat{v}^2 Pe$ are neglected in the series then the concentration expression is identical to the adsorption (with errors of the order 7\%). Retaining the first order term will give errors of the order 0.04\%. So the small difference between the form of the concentration curves and adsorption, discussed in \S \ref{ResSec}, can now be seen as an effect of diffusion in the gas.

The position of the front of the concentration wave is
\begin{align}
s(t) & = \widehat{v} {\cal L} k_1 t + s_0
=  \frac{\epsilon u C_0}{(1-\epsilon)  \rho_q \tilde{q}^* + \epsilon C_0 } t + s_0
\label{stTW}
\, ,
\end{align}
where $s_0$ is a constant to be determined.


\subsection{Using the travelling wave to characterise experiments}

Since measurements are primarily taken at the outlet we may use the above expressions to characterise breakthrough.
If we denote
$\Lambda =\exp(((L-s_0)/\hat{v}{\cal L})$ then the concentration and amount of adsorbate at the outlet are described by
\begin{align}
\label{CqTW}
\widetilde{C}(L,t)
& \approx  C_0 \left( 1 -  \alpha \Lambda e^{-k_1 t} \right) ~ ,  \qquad
\tilde{q}(L,t)  = \tilde{q}^* \left( 1 -\Lambda e^{-k_1 t}\right) \, .
\end{align}
Note, in both the adsorption and absorption literature there exist a number of models to predict the shape of the breakthrough curve based on assumptions regarding the rate of decline of CO$_2$ at the exit, typically that it is proportional to the amount of CO$_2$ remaining in the gas. These lead to a classic logistic type equation which integrates to an exponential relation between the concentration and time. For example the Thomas and Yoon-Nelson models take identical forms
\bea
\widetilde{C}(L,t)
=  \frac{C_0}{1+A e^{-k_1 t} }  \, ,
\eea
with different definitions for the constants, see \cite{Lin99,Han09}. Assuming the exponential term to be small, the first order Taylor series leads to the form of equation (\ref{CqTW}a). Below we will see that $\Lambda \sim A$ is large hence these earlier models can provide a good approximation to the later stages of breakthrough (when $A e^{-k_1t} \ll 1$). Further, since the physics of the process is neglected they do not show the dependence on the problem parameters, nor do they show any other quantities such as the amount adsorbed or the evolution of $s$.

We may use the breakthrough curve to determine the values of $k_1, \Lambda$ without worrying about $s_0$.
Consider two points on the experimental breakthrough curve (here we choose them  at either side of the central region). From Figure \ref{fig:BT} we find $\widetilde{C}(L,t_1)  = 0.164 C_0$ when $t_1 = 60\times  11.16$\,s and $\widetilde{C}(L,t_2) = 0.828 C_0$ at $t_2 = 60 \times 13.17$\,s.
This allows us to write down
\bea
\frac{\widetilde{C}(L,t_1)}{C_0} = 0.164 =    1 -  \alpha \Lambda e^{-k_1 t_1}  \qquad ~ ,\qquad \frac{\widetilde{C}(L,t_2)}{C_0} =0.838 =    1 - \alpha \Lambda e^{-k_1 t_2} \, .
\eea
Eliminating $\Lambda$ between the two equations leads to
\bea
k_1 = \frac{1}{t_2-t_1} \ln \left(\frac{C_0-\widetilde{C}(L,t_1)}{C_0- \widetilde{C}(L,t_2)}\right) \approx 0.0136 {\mbox ~ s}^{-1}\, .
\eea
This process was repeated using a number of data points within the breakthrough curve, to determine an average $k_1 = 0.0137$\,s$^{-1}$.
This value is within the typical range reported in \cite{Shaf15} and has been used in our numerical calculations, see Table \ref{tab:Table1}.
Taking this value of $k_1$ the unknown $\Lambda$ may be found,
\bea
\Lambda = \left(1-\frac{\widetilde{C}(L,t_1)}{C_0} \right) e^{k_1 t_1}= \left(1-\frac{\widetilde{C}(L,t_2)}{C_0} \right) e^{k_1 t_2}  \approx 8050 \, .
\eea
The concentration and amount of adsorbate at the outlet may therefore be determined without knowledge of $s_0$.

In \S \ref{NonDimSec} we derived an exact solution for the amount of adsorbate, equation \eqref{qanal}. At the end of the column the dimensional version of this solution is
\begin{align}
\widetilde{q}(x,t) & = \tilde{q}^* \left( 1 -  e^{-k_1( t-t_{s}(L))}\right) \, .
\end{align}
Comparison with equation \eqref{CqTW}) shows that the travelling wave reproduces the exact solution.

\begin{figure}
\centering
\includegraphics[width=0.5\textwidth]{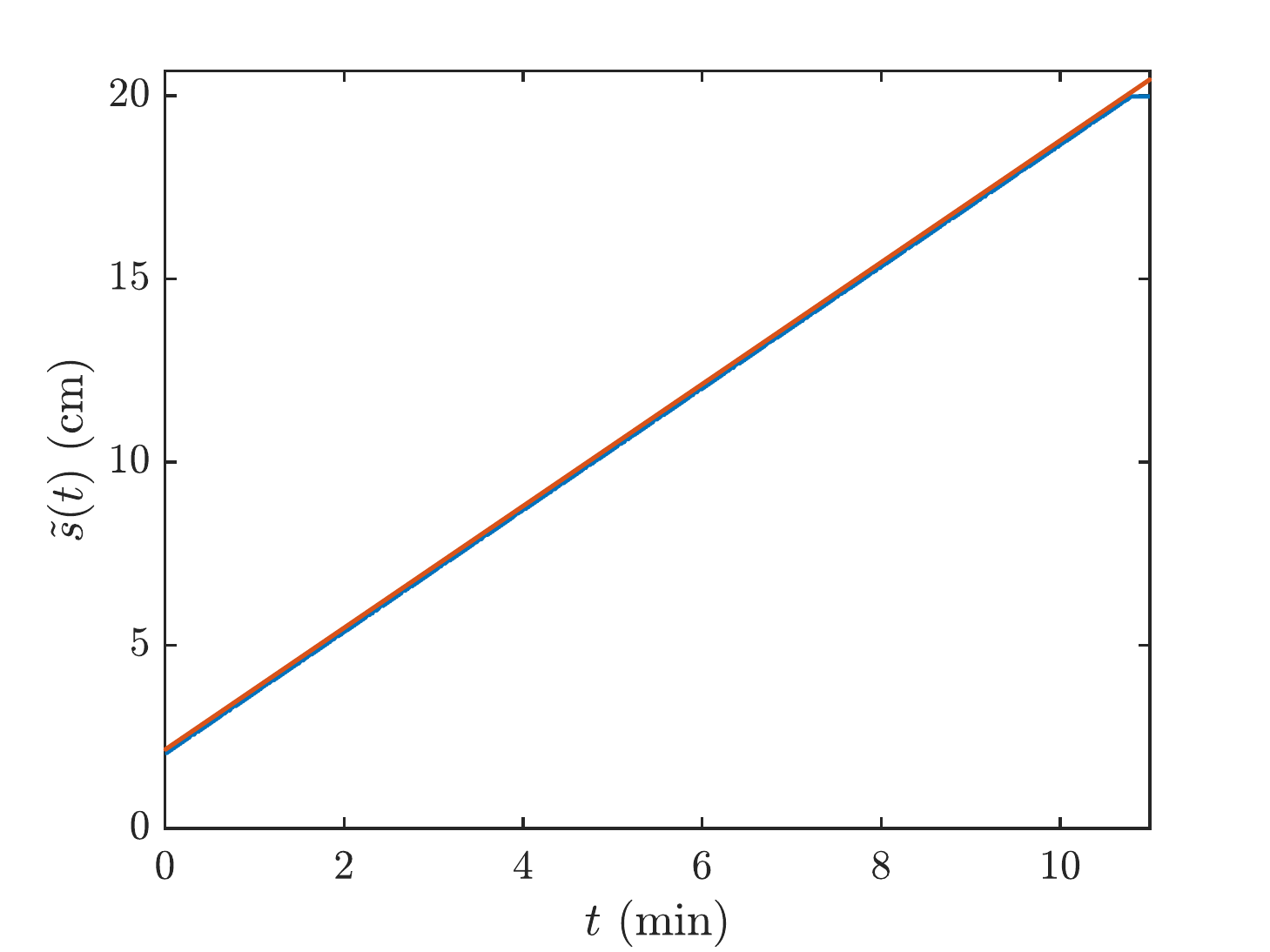}
\caption{Evolution of the point, $\tilde{s}(t)$, where the concentration of CO$_2$ reaches zero.
}
\label{fig:st}
\end{figure}

The unknown $s_0$ may be determined in a number of ways, for example we could look at the time of first breakthrough predicted by the numerical solution, in this case $t_{fb} \approx 10.75$ minutes. Then, according to \eqref{stTW}, we obtain $s_0 = L- \widehat{v} {\cal L} k_1 t_{fb} \approx 0.0215$m. We could also use the definition  of $\Lambda$ to obtain
$
s_0 \approx L - {\cal L} \ln \Lambda \approx 0.0201$ m.
However, since this definition  involves large exponentials it is a sensitive quantity. In Figure \ref{fig:st} we plot the evolution of the front of the concentration wave predicted by the numerical scheme and also the linear approximation of equation \eqref{stTW} with $s_0=0.0215$\,m. The numerical solution has the short flat section at the end. The flat section of the numerical result merely indicates that the gas has reached the outlet and the front is no longer being tracked. Obviously the agreement is excellent. Equation \eqref{stTW} can therefore be used  to calculate the breakthrough time under different conditions, i.e. longer columns, different flow rates or different initial concentration.

An analytical expression for $t_{fb}$ can be written  in terms of the system parameters by setting $s=L$ in equation \eqref{stTW}
\begin{align}
\label{tfbeq}
t_{fb}  & =  \frac{L-s_0}{\hat{v} {\cal L} k_1} = \frac{(1-\epsilon)  \rho_{q} \tilde{q}^* + \epsilon C_0} {\epsilon u C_0}  (L-s_0) \approx  \frac{(1-\epsilon)  \rho_{q} \tilde{q}^*} {\epsilon u C_0}  (L-s_0)
\, .
\end{align}
The term $(1-\epsilon)  \rho_{q} \tilde{q}^*$ is related to the mass of carbon adsorbed during the experiment, whereas the term
$\epsilon C_0$ represents the mass of gaseous CO$_2$ in the column. This second term is about 1\% of the first and so may be  neglected. For standard operating conditions, $s_0 \ll L$, so we may also neglect $s_0$. Equation \eqref{tfbeq} then makes clear all the factors which affect the breakthrough time and, importantly, it is independent of $k_1$.

If we define the process time as being when the outlet concentration is 90\% of its inlet value then by setting $\widetilde{C}/C_0 = 0.9$ in  equation \eqref{CqTW} we find
\bea
t_{90}=t_{fb}+ \frac{1}{k_1} \log(10 \alpha) \, ,
\eea
where we have made use of the fact $\log \Lambda = (L-s_0)/(\widehat{v} {\cal L}) = k_1 t_{fb}$ to reduce this expression.
The final term  represents the time taken for the width of the wave to pass through the outlet.

Previously we have discussed the choice of the adsorbate density. If we obtain a breakthrough time from the experimental data then we may use this to calculate
\begin{align}
\label{rhoqeq}
\rho_{q} = \frac{\epsilon C_0}{(1-\epsilon)  \tilde{q}^*}\left[ \frac{u t_{fb}}{L-s_0}-1 \right] \approx \frac{\epsilon C_0}{(1-\epsilon)  \tilde{q}^*}\left( \frac{u t_{fb}}{L-s_0} \right)
\, .
\end{align}
The density of adsorbate is discussed in \cite{Baha}, between 1 and 5 bar it varies  between approximately 200 and 400 kg/m$^3$. The data of \cite{Shaf15} shows that breakthrough occurs somewhere between 10.17 and 10.49 minutes. Taking the average $t_{fb} = 60 \times 10.33$\,s equation \eqref{rhoqeq} indicates $\rho_{q} = 311.3$\,kg/m$^3$. Given that there are a number of approximations involved in this work as well as experimental uncertainty (such as not knowing the exact time of first breakthrough) we take this value as an initial guess.
In Figure \ref{fig:BT} we adjust it by 4.5\%, $\rho_{q} = 325$ kg/m$^3$, to obtain  the good agreement with the breakthrough curve. This is the only fitting of our model.

Of course there are other ways to calculate the density of the adsorbate. For example, the mass of adsorbed carbon can easily be measured from the experiment by comparing the starting mass of the column with that at the end. The mass flux of CO$_2$ entering the column is $  M_{CO2}C_0 u$ kg/m$^2$\,s (where $M_{CO2} = 44.01 \times 10^{-3}$\,kg/mol is the molar mass of CO$_2$) this occurs over an area $\epsilon \pi R^2$ m$^2$. Hence the mass of CO$_2$ in the column at any time may be expressed as
\bea
M_{ex} =\left\{
\begin{array}{ll}
M_{CO2}u  \epsilon \pi R^2 C_0 ~ t & t \le t_{fb}\\
M_{CO2}u  \epsilon \pi R^2 ~ \left(C_0t_{fb} + \int_{t_{fb}}^t C_0 - \widetilde{C}(L,t) ~ dt \right) & t \ge t_{fb} ~ .
\end{array}
\right.
\label{Moptions}
\eea
The final integral is easily evaluated to give
\bea
\label{Mtgfb}
M_{ex} = M_{CO2}u  \epsilon \pi R^2C_0 \left[t_{fb} + \frac{\alpha \Lambda}{k_1}  \left(\exp(-k_1 t_{fb}) - \exp(-k_1 t) \right)\right] \qquad  t \ge t_{fb} ~ ,
\eea
where $t_{fb}$ is given by equation \eqref{tfbeq}.
These formulae demonstrate that before first breakthrough the increase in mass is linear in time, afterwards there is an exponential decrease in the rate that mass is added. In the limit of large time, $t \ra \infty$, this expression reduces to
\bea
M_{lim} \ra  M_{CO2}u  \epsilon \pi R^2C_0 \left[t_{fb} + \frac{\alpha }{k_1}  \right] \,  \, .
\eea
This is the maximum mass that may be extracted during the experiment.

To determine the density we could simply stop an experiment at first breakthrough, determine the adsorbed mass and then from equations \eqref{tfbeq}, \eqref{Moptions} we obtain
\bea
\label{Mfbeq}
\rho_{q}= \frac{1}{(1-\epsilon)\tilde{q}^*} \left(\frac{M_{fb}}{M_{CO2} \pi R^2(L-s_0)} - \epsilon C_0\right)    ~ .
\eea
The final term in the  brackets represents the mass of the gas in the column: this is approximately 1\% of the total mass, hence
\bea
\label{Mfbeqapp}
\rho_{q} \approx   \frac{M_{fb}}{M_{CO2} \pi R^2(1-\epsilon)\tilde{q}^*(L-s_0)}  \qquad {\mbox or} \qquad
M_{fb} \approx  \rho_{q}  M_{CO2} \pi R^2(1-\epsilon)\tilde{q}^*(L-s_0)
~ .
\eea
The first expression gives a simple formula for the density of adsorbate  in terms of the system parameters and the mass of adsorbed CO$_2$ at first breakthrough. The second expression shows how the mass at first breakthrough depends on the syetem parameters. So, to increase the adsorbed mass requires either an increase in column dimensions, adsorbate density or equilibrium concentration or a decrease in $\epsilon$. The interstitial velocity and adsorption rate constant \emph{play no role in the mass adsorbed at first breakthrough}. However the interstitial velocity does affect the time for first breakthrough, as evidenced by equation \eqref{tfbeq}.
If the measurement is made some time after first breakthrough then  equation \eqref{Mtgfb} does show a dependence on $u$ and $k_1$, however even in the limit $t \ra \infty$ this contribution only accounts for 10\% of the adsorbed mass, so the dependence on these two parameters is weak.

\begin{figure}
\centering
\includegraphics[width=0.5\textwidth]{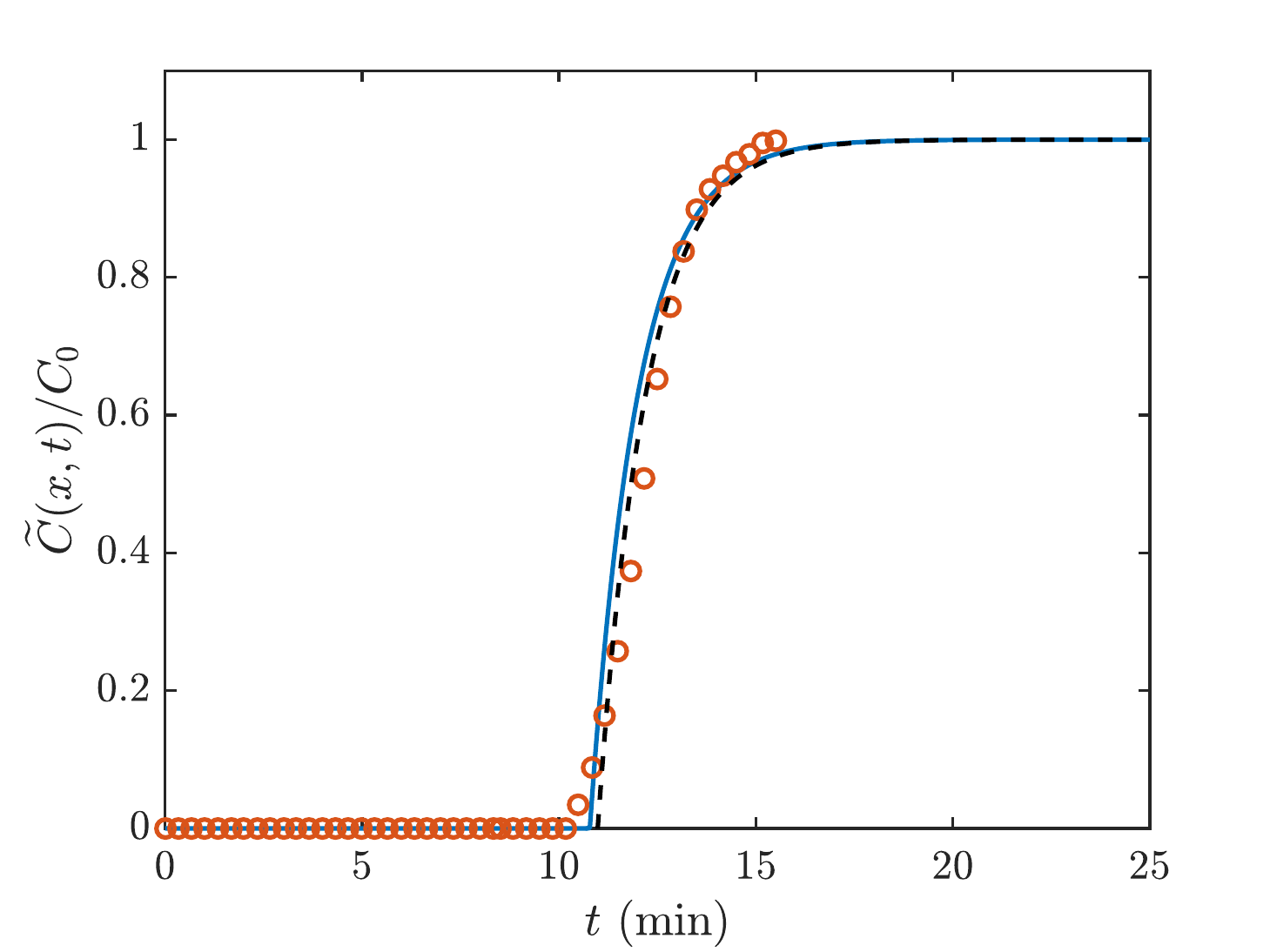}
\caption{Comparison of the travelling wave solution for the gas concentration (dashed line) against the numerical solution (solid line) and experiment (circles). }
\label{fig:TW}
\end{figure}

The approximate expression for $\widetilde{C}$, equation \eqref{CqTW}, is compared with the experimental data for breakthrough and also the numerical solution in Figure \ref{fig:TW} (using $\rho_{q} = 325$). The full expression for the travelling wave  {(the first part of equation (34) in the supplementary information)} is not plotted because it is indistinguishable from the approximate version.
Clearly the approximate expression for  the travelling wave provides excellent agreement with the experimental data. So we may use it to characterise breakthrough.

To find the width of the reaction zone
consider the equation for $\widetilde{C}(x,t)$, (\ref{Cdim}a). If we define the reaction zone as having $\widetilde{C}(x,t) \in [0,0.9]C_0$ then the co-ordinates where the extreme values are achieved at a given time $t$ are
\begin{align}
\left. x \right|_{\widetilde{C}=0} &= s_0 + \widehat{v}{\cal L}\left[k_1 t + \log\left(\frac{\widehat{v}^2 Pe  -1}{\widehat{v}^2 Pe}\right)\right]\,,
\\
\left. x \right|_{\widetilde{C}=0.9C_0} &=  s_0 + \widehat{v}{\cal L}\left[k_1 t + \log\left(0.1 \frac{\widehat{v}^2 Pe  -1}{\widehat{v}^2 Pe}\right)\right] \, .
\end{align}
The width of the reaction zone is then
\begin{align}
\label{Weq}
W = \left. x \right|_{\widetilde{C}=0} -
\left. x \right|_{\widetilde{C}=0.9C_0} &= - \widehat{v}{\cal L} \ln 0.1  \approx   \frac{\epsilon u C_0}{(1-\epsilon) k_1 \rho_{q} \tilde{q}^* } \ln 10 \, .
\end{align}
To verify this we take the values
provided in Table \ref{tab:Table1} and obtain $W \approx 2.3  {\cal L} \approx 4.7$cm. The corresponding width from the numerical solution is  $2.337{\cal L} \approx 4.8$cm (less than 3\% difference).
A formula for the width of the reaction zone is provided in \cite{Zhao}. If we define this using the times for first breakthrough and when $\widetilde{C}(x,t)/C_0 =0.9$ then
\bea
W_{Z}= 2 L \frac{t_{0.9}-t_{fb}}{t_{0.9}+t_{fb}} \approx 5.2 {\mbox\, \text{cm}}\, .
\eea
This equates to  an 8\% increase from the numerical width. So this formula is slightly less accurate than  the present one, equation \eqref{Weq}. Equation \eqref{Weq} has the added advantage of clearly showing  the factors affecting the reaction zone.


\section{Conclusion}

The primary purpose of developing a mathematical model of a process is to help understand and so improve or optimise that process. Consequently, it is essential to start from the correct model. In this paper we have revisited the averaging process to derive an accurate system of equations describing the adsorption of a gas in a circular cross section packed column. In doing so we were able to identify a number of common errors in carbon capture which are prevalent the literature, such as:
\begin{enumerate}
\item the use of an incorrect adsorbate density (which then affects the mass sink and heat generation terms);
\item inconsistent neglect and retention of certain terms, such as heat conduction in the gas but not the solid, or heat transfer at the walls;
\item simultaneous inclusion of varying and constant velocity;
\item radial flow in previously radially averaged equations;
\item the incorrect integration of the adsorbate equation.
\end{enumerate}
Despite these errors many of the published works show excellent agreement with experimental data, often better than presented here. This may be explained through the choice of parameter values, for example, if an incorrect adsorbate  density is used it may be compensated for by adjusting the reaction rate. In general the models involve advection-diffusion equations, the experiment involves advection and diffusion so it is relatively straightforward to achieve good agreement using well chosen parameter values. However, the parameter values obtained will be incorrect and so lead to inaccurate predictions when the experiment is scaled up or altered in some way.

Once the correct averaged model was derived we proceeded to non-dimensionalise the system. This
showed which terms were important and which negligible. For example, it is clear  that diffusion of concentrate in the gas is negligible, hence a detailed calculation of its value is not necessary. Similarly conduction in the gas and solid are both small and have little effect on heat flow.

Numerical results were presented for an isothermal system. Good agreement was shown with experimental data. A brief analysis indicated that neglecting temperature variation, specifically on the adsorption saturation may lead to errors of the  same order as incurred by the neglect of mass loss, so future work should account for both of these effects.

The travelling wave solution removes the need for a numerical solution and provides a simple way to characterise the breakthrough curve. It also clearly demonstrates how the physical process depends on the operating conditions.
Specifically through this analysis we obtained exact expressions for
\begin{enumerate}
\item the time of first breakthrough;
\item the process time;
\item the width of the reaction zone;
\item the (time-dependent) mass of adsorbed CO$_2$ in the column.
\end{enumerate}

Perhaps the key parameter in all of this analysis is the adsorbed mass. The analytical solution shows that the main ways to increase this are by
increasing the adsorbate density, the column dimensions (both length and radius) and the adsorption saturation or by  decreasing the void fraction.
It is independent of the adsorption rate, gas velocity and diffusion.

The analysis in this paper has therefore led to an improved set of equations describing the adsorption of gas in a packed column. The novel analytical solution provides the relation between experimental operating conditions and outputs,  particularly the amount of adsorbed gas. The method is applicable to standard operating conditions for carbon capture in a column, as studied in numerous other papers.
Possible sources of error come from the neglect of temperature variation and mass loss. Whilst common practice to neglect these effects, temperature variation mainly affects the saturation concentration, while mass loss affects the fluid velocity. Heat is dissipated quickly (in comparison to the process time-scale) so this error will be limited by the specific temperature-saturation concentration relation and the heat diffusion rate. In both cases we expect their neglect to lead to errors of only a few percent. Consequently we do not expect either effect to be significant in the analysis but they are certainly worth investigating in future studies.

Although the theory  provides a clear description of the process, in practice it may not be sensible to alter certain parameters. For example the adsorption saturation increases with pressure and decreases with temperature, but there are operating costs associated with this. Decreasing void fraction increases the surface area for adsorption but also requires a higher pressure drop across the column to maintain the flow rate. A balance must therefore be found between the optimal theoretical solution and operating considerations.

\appendix
\renewcommand{\thesubsection}{\Alph{subsection}}

\section{Isotherm equation}\label{qSec}

\begin{table}[h]
\centering
\begin{tabular}{ccccccc}
\hline
& $q_{m0}$ (mol/kg) & $K_{0}$ (1/atm) & $n_0$ & $\alpha_T$ & ${\eta}_T$ & $\Delta H$  (kJ/mol) \\
\hline
Physical adsorption&  3.57 & 0.66 &  0.65 & 1.05 & 12.45 & 22.23 \\
Chemical adsorption & 0.69 & 8.14 $\times 10^4$ & 0.27 & 0.22 & 2.62 & 73.24\\
\hline
\end{tabular}
\caption{Isotherm parameters from  \cite[Table 2]{Shaf15}.}
\label{TableToth}
\end{table}

In \cite{Shaf15} the adsorption saturation is assumed to consist of a chemical and physical component
\bea
\tilde{q}^*= q_c + q_p
\eea
where each component takes the form
\bea
q = \frac{q_m K_T P}{(1+(K_T P)^{n})^{1/n}} \, .
\eea
Certain parameters show a pressure and temperature dependence
\begin{align}
K_T = K_0 \exp\left[\frac{\Delta H}{R T_0}\left(\frac{T_0-T}{T}\right)\right] & \qquad q_m = q_{m0} \exp\left[{\eta_T}\left(\frac{T_0-T}{T_0}\right)\right] \\
n& =n_0 + \alpha_T \left(\frac{T-T_0}{T}\right) \, .
\end{align}
The values for the various constants are shown in Table \ref{TableToth}. In all of our calculations the reference temperature $T_0$ was set to 303\,K, which corresponds to the base temperature in the considered experiment.

\section*{Acknowledgements}

T. G. Myers acknowledges financial support from the Ministerio de Ciencia e Innovación, Spain Grant No. MTM2017-82317-P. F. Font acknowledges that the research leading to these results
has received funding from la Caixa Foundation and MICINN through the Juan de la Cierva programme, Grant No. IJC2018-038463-I.

\bibliographystyle{plain}
\bibliography{bibcarbon}

\end{document}